\documentclass[preprint2]{aastex}
\newcommand{\REV}[1]{{\bf #1}}         % turn on highlighting of revisions.
\renewcommand{\REV}[1]{#1}             % this turns highlighting off.

\newcommand{\vwc} {VW~Cep}
\newcommand{\ffb} {44~Boo}
\newcommand{\aped} {APED}
\newcommand{\asca} {{\it ASCA}}
\newcommand{\chan} {{\it Chandra}}
\newcommand{\xmm}  {{\it XMM-Newton}}
\newcommand{\cxo} {CXO}
\newcommand{\ciao} {CIAO}
\newcommand{\dem} {\mbox{$DEM$}}
\newcommand{\heg}  {HEG}
\newcommand{\hetgs} {HETGS}
\newcommand{\letgs} {LETGS}
\newcommand{\meg}  {MEG}

\newcounter{ion}
\newcommand{\eli}[2]{\setcounter{ion}{#2}#1{~\sc\roman{ion}}}

\newcommand{\mone}{^{-1}}
\newcommand{\mtwo}{^{-2}}
\newcommand{\mthree}{^{-3}}

\newcommand{\kms}{\mathrm{\,km\,s\mone}}

\slugcomment{Rev: \today}

\shorttitle{X-ray Spectroscopy of the Contact Binary VW Cephei}
\shortauthors{Huenemoerder, et. al}

\begin{document}

\title{X-ray Spectroscopy of the Contact Binary VW Cephei\footnote{23 June 2006: Accepted for publication in The Astrophysical Journal}}

\author{David P. Huenemoerder, Paola Testa }
\affil{Kavli Institute for Astrophysics and Space Research\\
  Massachusetts Institute of Technology \\
  Cambridge, MA 02139}
\email{dph@space.mit.edu, testa@space.mit.edu}

\and

\author{Derek L.\ Buzasi}
\affil{US Air Force Academy \\
  Dept.\ of Physics,\\
  HQ USAFA/DFP\\
  2354 Fairchild Dr., Ste.\ 2A31\\
  USAF Academy, CO 80840
}
\email{Derek.Buzasi@usafa.af.mil}

\begin{abstract}

  Short-period binaries represent extreme cases in the generation of
  stellar coronae via a rotational dynamo.  Such stars are important
  for probing the origin and nature of coronae in the regimes of rapid
  rotation and activity saturation. VW~Cep ($P=0.28$~d) is a relatively
  bright, partially eclipsing, and very active object.  Light curves
  made from \chan/\hetgs\ data show flaring and rotational modulation,
  but no eclipses.  Velocity modulation of emission lines
  indicates that one component dominates the X-ray emission.  The
  emission measure is highly structured, having three peaks.
  Helium-like triplet lines give electron densities of about
  $3-18\times10^{10}\, \mathrm{cm\mthree}$. We conclude that the
  corona is predominantly on the polar regions of the primary star and
  compact. 

\end{abstract}


\section{Introduction}

\object{VW~Cep} (\object{HD~197433}), one of the X-ray-brightest of
contact binaries, is a W-type W~UMa system --- one in which the more
massive and larger star has {\em lower} mean surface brightness such
that the deeper photometric eclipse occurs during the occultation of
the smaller star.  VW~Cep has an 0.28 d (24 ks) orbital period, is partially
eclipsing ($i=63^\circ$), and has component spectral types of K0~V and
G5~V \citep{Hill:1989,Hendry:Mochnacki:2000}, or, according to
\citet{Khajavi:al:2002}, F5 and G0.

Among the coronally active binaries, activity is strongly correlated
with the rotation rate, and in particular, the Rossby number, which is
a measure of the relative importance of Coriolis forces in the
convective layer, which is in turn related to the magnetic dynamo
strength \citep{Pallavicini:1989}. At periods below one day, activity
saturates \citep{Vilhu:Rucinski:1983,Cruddace:Dupree:1984}.  Since the
activity level, as defined by $L_x/L_\mathrm{bol}$, actually decreases
with increasing rotation rate, this trend has been referred to as
``super-saturation'' \citep{Prosser:Randich:al:1996,
  Randich:1998,Jardine:Unruh:1999,
  James:Jardine:al:2000,Stepien:Schmitt:Voges:2001}.  
\REV{The saturation level of $\log(L_x/L_\mathrm{bol})\sim -3$
  extends down to periods of 0.4 days.  For contact binaries with
  periods between $0.2$ to $0.4$ days, the median $L_x/L_\mathrm{bol}$
  is lower by a factor of four.}
\citet{Buzasi:1997} showed that in rapidly rotating low-mass stars,
magnetic loops would be swept to the poles.  Super-saturation may
occur because loops are large and are unstable to the Coriolis forces
as they exceed the co-rotation radius: extended loops get swept to the
poles \citep{Jardine:Unruh:1999}.  Or loops could be compact (relative
to the stellar radius), and the dynamics of surface flows clear
equatorial regions \citep{Stepien:Schmitt:Voges:2001}.  The two
scenarios are similar in that activity is predominantly polar, but
they differ in an important respect: X-ray sources are either large
volume and rarefied or low volume and dense.  Assuming correlation
between photospheric spots and coronal emission, optical light curve
modeling is consistent with either scenario: Doppler image maps of
VW~Cep \citep{Hendry:Mochnacki:2000} showed large polar spots.  X-ray
light curve modeling has been difficult because of the high
probability of confusion by flaring, lack of phase redundancy, and the
unavailability of high-resolution spectral diagnostics.

Previous high-energy observations have characterized the coronae of VW
Cep.  \cite{Choi:Dotani:1998} analyzed \asca\ spectra, and found a
flux of about $1\times10^{-11}$ $\mathrm{ergs\,cm^{-2}\,s^{-1}}$, and
two component model temperatures of $7$ and $22\times10^6$K ($\log
T\sim 6.8$ and $7.3$) with about equal emission measures.  They
obtained significantly reduced abundances of Fe, Si, Mg, and O, but a
Solar value for Ne.  A flare also occurred during the observation,
with a factor of three increase in the count rate.  They used the
flare emission measure and loop-scaling models to estimate a density
of about $5\times10^{10}\mathrm{cm^{-3}}$.  The Ginga observations
\citep{Tsuru:al:1992} showed a thermal plasma temperature in excess
of $10^8\,$K, a flux similar to that determined by
\cite{Choi:Dotani:1998}, no rotational modulation, and Fe~K flux lower
than expected.

The value of $L_x/L_\mathrm{bol}\sim -3.6$ places VW~Cep in the
super-saturated regime (using the bolometric value implied by 
\citet{Stepien:Schmitt:Voges:2001}).

%\section{High Resolution X-Ray Spectra}\label{sec:hires}
\section{Observations and Reduction of High Resolution X-Ray Spectra}\label{sec:hires}

We observed VW~Cep for 116~ks with the \chan\ High Energy Transmission
Grating Spectrometer (\hetgs) on August 29-30, 2003 (observation
identifier 3766).  The instrument has a resolving power ($E/\Delta E$)
of up to 1000 and wavelength coverage from about 1.5~\AA\ to 26~\AA\
in two independent channels, the High Energy Grating (\heg), and
Medium Energy Grating (\meg). For more details on the \hetgs, see
\citet{HETG:2005}.

VW~Cep has also been observed in X-rays with the Chandra Low Energy
Transmission Grating Spectrometer \citep{Hoogerwerf:al:2003}, and with
the \xmm\ X-ray observatory \citep{Gondoin:2004a}.  The only other
contact binary observed with \hetgs\ is 44~Boo \citep{Brickhouse:01}.
Observations with the \hetgs\ provide the improved spectral resolution
and sensitivity required to better determine line fluxes, wavelengths,
and profiles.  Ultimately, we wish to determine which coronal
characteristics are truly dependent upon fundamental stellar
parameters.  Combination of VW~Cep data with that from \letgs\ and
\xmm, as well as with results for other short period systems (e.g.,
44~Boo, ER~Vul) will help us to understand the basic emission
mechanisms.

Data were calibrated with standard CIAO pipeline and response
tools\footnote{\tt http://cxc.harvard.edu/ciao} ({\tt CIAO 3.2
  Thursday, December 2, 2004 / ASCDS version number}).  The updated
calibration database geometry file\footnote{\tt
  telD1999-07-23geomN0005.fits} provided a significant improvement in
wavelength scales which resulted in good agreement between \heg\ and
\meg\ as well as between positive and negative diffraction orders.
This was important for line centroid analysis.

\begin{figure*}[ht]     
%  \plotone{vwc_spec_full.ps}
{\epsscale{2}    % pp
\plotone{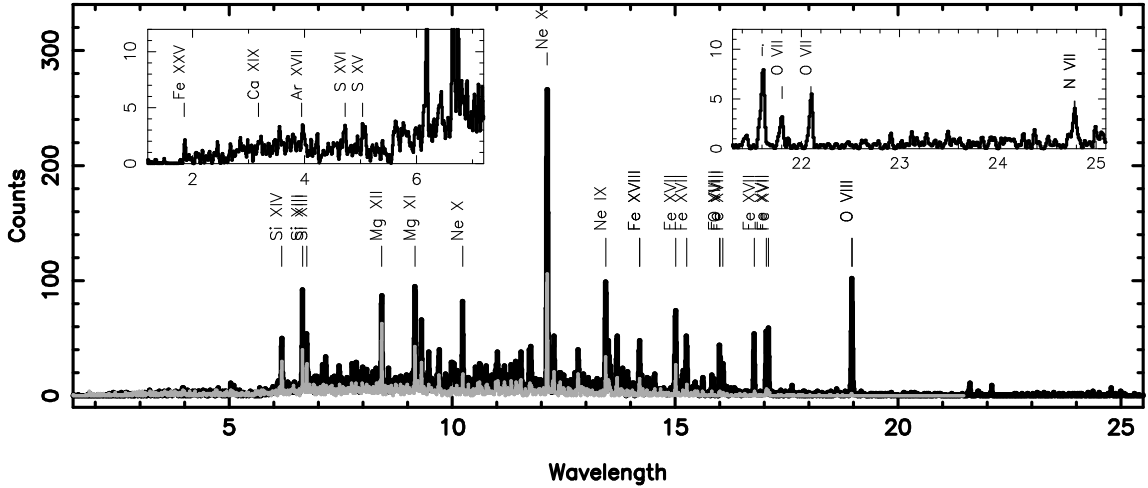}
}
%\centering\leavevmode\scalebox{1.15}{\includegraphics{vwc_spec_full.ps}}
  \caption{
    HETGS spectrum of VW~Cep, 116 ks exposure.  MEG (16433 counts) is
    the black line, and HEG (5269 counts) the gray.  Insets show
    detail of the HEG spectrum (left) and MEG O~VII triplet and N~VII
    region (right).
    \label{fig:spec}
  }
\end{figure*}

Line, emission measure, and temporal analyses were done in ISIS
\citep{Houck:00,Houck:2002} and with custom code written in the S-Lang
scripting language\footnote{S-Lang is available from {\tt
    http://www.s-lang.org/}} using ISIS as a development platform.
\notetoeditor{Figures 1,3,7,8 are black and white or gray-scale; all
  others have color.  Figures 1, 5, and 6 require 2 columns for
  adequate presentation.  All others can be formatted into one
  column. }
\placefigure{fig:spec}   %\input{figspec}
\notetoeditor{figure~\ref{fig:spec} should be two columns wide}
%
%\clearpage              % mod for preprint
%\clearpage
%
Figure~\ref{fig:spec} shows the counts spectrum obtained from the
\hetgs\ full exposure.  The spectrum is qualitatively typical of coronal
sources: a variety of emission lines from highly ionized elements
formed over a broad temperature region, from \ion{O}{7}, \ion{N}{7},
\ion{Ne}{9}, and \ion{Fe}{17} ($\log T\sim 6.3$--$6.7$), up to high
temperature species like \ion{S}{15}, \ion{S}{16}, \ion{Ca}{19},
and \ion{Fe}{25} ($\log T\sim 7.2$--$7.8$).  It is 
apparent that iron has a fairly low abundance relative to neon, given
the relative weakness of the 15\AA\ and 17\AA\ \ion{Fe}{17} lines
relative to the \ion{Ne}{9} 13\AA\ lines.  The observed flux in the
2--25~\AA\ is $8.4\times10^{-12}\,\mathrm{ergs\,cm\mtwo\,s\mone}$, and
the luminosity (for a distance of 27.65 pc \citep{Perryman:97a} is
$7.7\times10^{29}\,\mathrm{ergs\,s\mone}$.

\section{Analysis}

\subsection{Light and Phase Curves}
\begin{figure}
%  \epsscale{0.8}
%    \plotone{vwc_tglc.ps}
  \epsscale{1.0}    % mod for preprint
\plotone{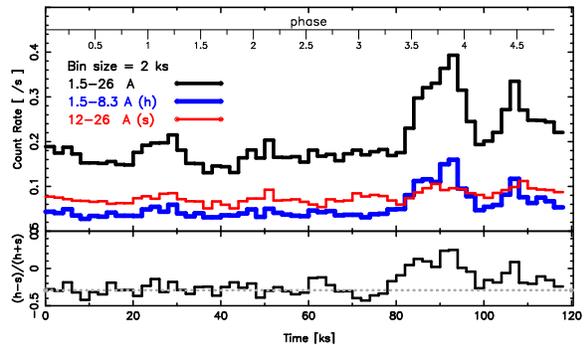}
    \caption{Count rates of VW~Cep in 2 ks bins.  In the upper panel
      are light curves extracted from all diffracted photons
      (1.5-26\AA; upper heavy curve), a ``soft'' band (12-26\AA;
      lower, thinner curve), and a ``hard'' band (1.5-8.3\AA; lower,
      thicker curve (blue)).  The lower panel shows a hardness ratio
      (solid histogram), and the median of the hardness for the first
      70 ks (light dotted line).  We defined the non-flare state as
      the first 80 ks.
      \REV{At the top we show an axis in which the
        integer part gives the number of rotations and the fractional
        part is the phase.}
    }
    \label{fig:lc}
\end{figure}

\begin{figure}[h]
%  \plotone{vwc_tgpc.ps}
\epsscale{1.0} % mod for preprint
  \plotone{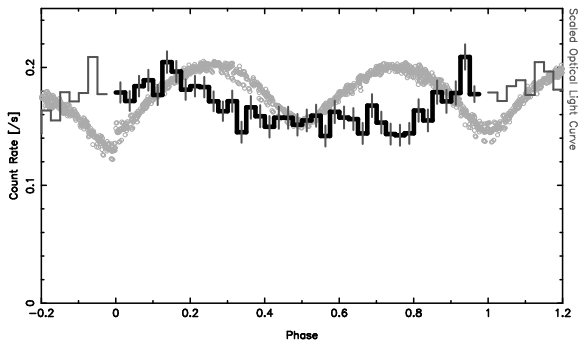}
  \caption{ The dark, solid histogram is the phase-folded X-ray
    count-rate curve of VW~Cep, excluding flare times.  Bins are
    $\Delta\phi=0.025$, or 600 s, with a cumulative exposure per bin
    of 1800 to 2400 s, depending on the phase.  
\REV{Errorbars on the  histogram are the $1\sigma$ statistical uncertainties.}
    The gray points are a scaled visual
    intensity light curve to show the very different character of the
    optical and X-ray light spatial distributions (Visual photometric
    data are with permission from T.\ Pribulla at {\tt
      http://www.astro.sk/$\sim$pribulla/lc/vwcep.dat}.)
    \label{fig:pc}
  }
\end{figure}

\placefigure{fig:lc}  %\input{figlc}
%\clearpage
%\clearpage
%
The exposure lasted for five revolutions without interruption.  Such
phase redundancy is important to discriminate intrinsic variability
from that caused by rotational or eclipse modulation (for example, see
the UV and X-ray monitoring studies of AR\ Lac,
\citet{Neff:89,Huenemoerder:Canizares:al:2003}).  Figure~\ref{fig:lc}
shows light curves\footnote{\raggedright Light and phase curves were
  made with custom software (the {\tt aglc} S-Lang package) available
  on the Chandra Contributed Software site: {\tt
    http://cxc.harvard.edu/cont-soft/software/aglc.1.2.3.html}} of
diffracted photons (\heg\ and \meg, orders $-3$ to $3$, excluding
zeroth) covering the entire \hetgs\ spectrum as well as for two wide
bands covering short 
\REV{(``hard'', 1.5--8.3\AA) and long (``soft'', 12--26\AA)}
wavelength regions.  The hardness counts ratio (defined as
$(hard-soft)/(hard+soft)$) shows that the large increase after 80 ks
is
\REV{ probably a flare: proportionally more flux was emitted at
  shorter wavelengths, which are very sensitive to high temperatures
  through thermal continuum emission.  Conversely, the bump in count
  rate between 20-30 ks does not show in hardness, and likely is due
  to rotational modulation.

  We assume that flares are hot and will thus show a change in
  hardness, and that changes in volume alone will primarily show a
  change in count rate.  More complicated scenarios are possible, such
  as extended loops with temperature gradients rotating in and out of
  view and causing both hardness and count-rate variability on the
  time scale of a rotation.  However, we know that flares --- defined
  as hot, impulsive events with a gradual decay, in analogy to flares
  resolved on the Sun --- are frequent on most active stars, and we
  will adopt the assumption as our working hypothesis.  }
\placefigure{fig:pc}   %\input{figpc}
%
%\clearpage
%\clearpage
%
Rotational modulation is apparent in the phase-folded curve of the
non-flare times, which covers about three orbital periods.
Figure~\ref{fig:pc} shows this curve, using the ephemeris of
\cite{Pribulla:al:2002}.  No eclipses are obvious 
\REV{but there is gradual modulation on the orbital period with an
  amplitude of 20\%.  The difference between the maximum and minimum,
  given the $1\sigma$ statistical errors shown, is about $5\sigma$.  }
The visual light curve \citep{Pribulla:al:2000} has a similar
amplitude but is very different qualitatively, having strong minima at
phases 0.0 and 0.5, and continuous variability in between (a trademark
of W~UMa systems).  The optical phase modulation is primarily
dependent upon the system geometry and inclination.

Without additional information, the X-ray light curve is not
sufficient for localizing emission to one star or another.  All we can
say is that there is some asymmetric distribution, and possible
occultation of longitudinally extended structures.  It does not appear
that there is significant emission from the smaller star, since there
is no secondary eclipse at phase 0.0. The X-ray light has a
distribution very different from the optical light.

\subsection{Velocity Modulation}

\REV{ In the high-resolution spectrum, additional information is
  available in the line positions which can help localize the emission
  by means of a determination of the mean radial velocity of the
  emitting plasma.  In any single feature, even the strongest,
  \ion{Ne}{10} (12.1 \AA), the line position is poorly constrained in
  phase bins small enough to sample the orbital velocity.
  \citet{Huenemoerder:Hunacek:2005}\footnote{Preprint:
    http://arxiv.org/abs/astro-ph/0409258} showed that the
  \ion{Ne}{10} mean Doppler velocity followed the primary's orbital
  radial velocity except for a sharp rise and fall across the
  secondary eclipse.  This ``flip-flop'' is known as the Rossiter
  effect \citep{Rossiter:1924} if it is due to resolution of the
  rotational velocity profile of the star through occultation of
  velocity ranges during a transit.  Given the low signal level
  available in a single line, this interpretation was not firm.  }

We have improved upon that measurement by accumulating signal from
many lines and constructing a composite line profile. This mixes
resolutions and line shape since long-wavelength lines have higher
velocity resolution than shorter wavelengths, but it does have the
advantage of increasing signal greatly.  Since HEG and MEG have
different resolutions and coverage, we kept their composites separate.
To avoid blending, we chose only fairly isolated features, and these
are flagged in the ``Use'' column of Table~\ref{tbl:linelist} with an
``H'' or an ``M''.

\begin{figure}[h]
%%  \includegraphics[scale=0.4]{CLP_meg.ps}
%  \includegraphics[scale=0.4]{f4a.eps}
%  \epsscale{0.4}
 % mod for preprint
  \centerline{\epsscale{1.0}\plotone{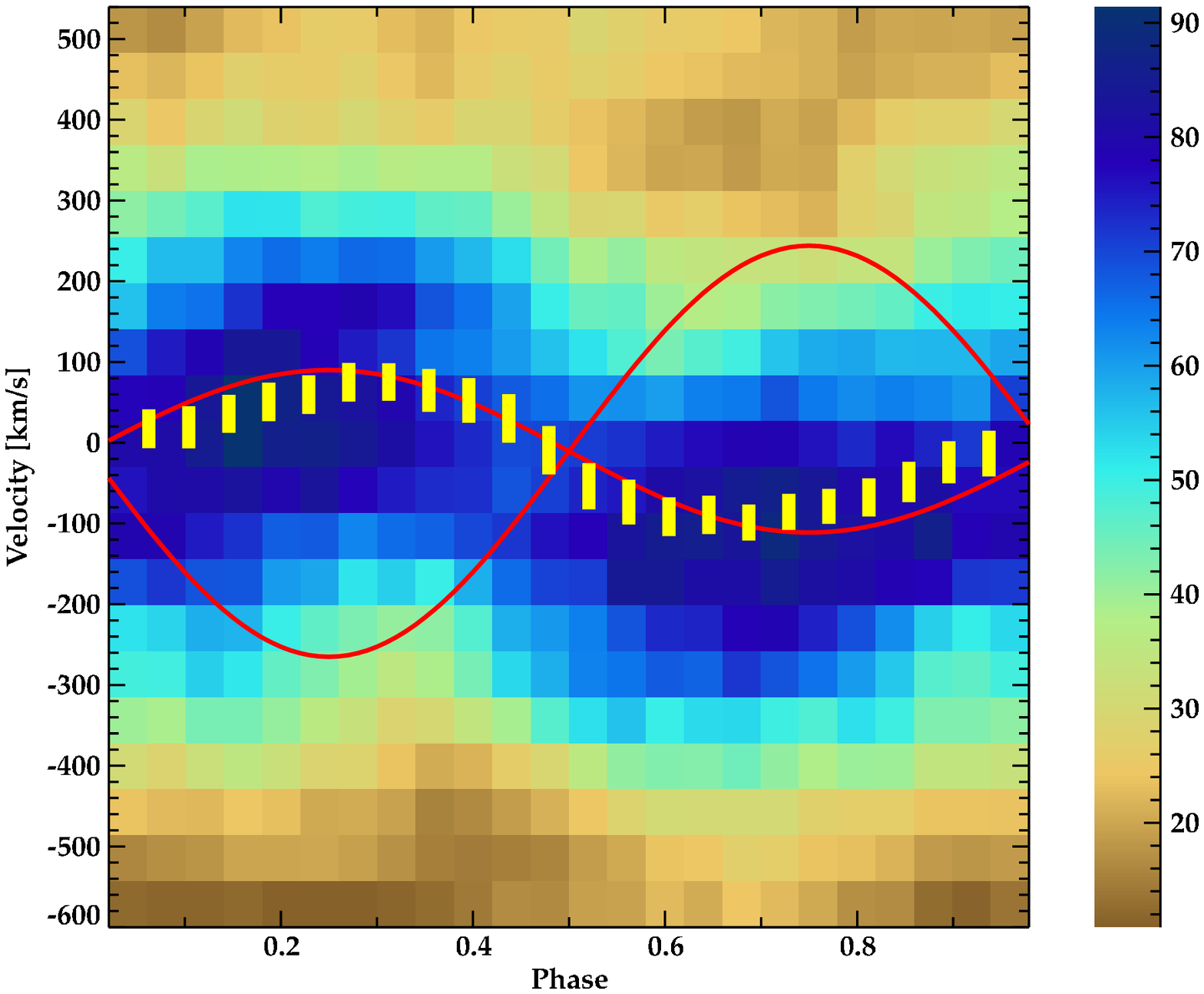}}
%  \newline
%  \includegraphics[scale=0.4]{f4b.eps}
  \centerline{\epsscale{1.0}\plotone{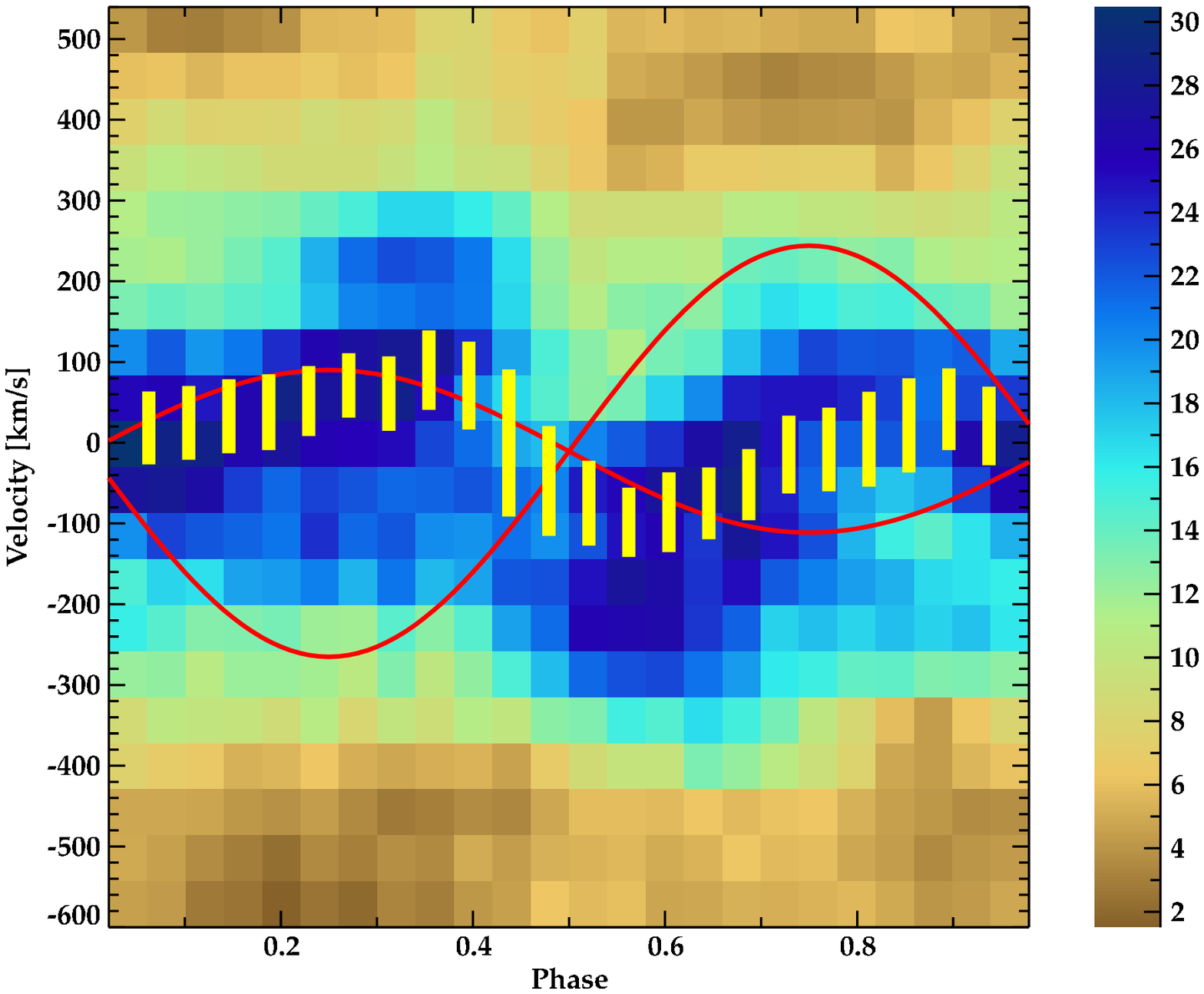}}
  \caption{ Composite line profile for the MEG spectrum (top) and HEG
    (bottom). Yellow bars are the 90\% confidence intervals of the
    centroid of the composite profile. The red sinusoidal curves are
    the center-of-mass velocities of the binary components.  The
    background is an intensity map of the composite profiles' counts.
    The bar to the right of each image gives the counts-to-color
    mapping for the image.  The systematic deviations from the
    primary's (more massive star) velocity are clear in the HEG for
    phases 0.5-1.0.  The centroids are correlated since they were
    measured in phase bins of width 0.2. 
 \REV{The red curves were
   also averaged over the same phase intervals as the data.
 }
    \label{fig:clp}
  }
\end{figure}

We accumulated spectra into 24 phase bins for each of the \heg\ and
\meg\ positive and negative first orders (96 distinct spectra).  We
then transformed about 15 lines in each spectrum from wavelength to a
common velocity scale and summed them.  Finally, we combined plus and
minus orders over a range of phase bins (to further improve the
signal) and fit Gaussians to the composite profiles' cores (defined to
be where the counts were greater than the maximum divided by $2e$)
separately for each of \heg\ and \meg.  This method is similar to that
used by \citet{Hoogerwerf:al:2004}.  We found that using five phase
bins (a running average over $\Delta\phi=0.2$) was adequate to provide
enough phase resolution and signal without losing too much sensitivity
to velocity changes.  
\REV{ To quantify the significance of the fitted velocity centroid, we
  computed the 90\% confidence intervals ($1.6\sigma$) for the
  velocity.  }
\placefigure{fig:clp}   %\input{figclp} 
\notetoeditor{figure~\ref{fig:clp} should be one column wide, w/ 2
  parts stacked vertically}
%
%\clearpage
%clearpage
Figure~\ref{fig:clp} shows the results. 
\REV{ The composite line profiles make the background image, with
  velocity on the $y$-axis, and phase on the $x$-axis, and the darker
  shading is for higher intensity as indicated by the color-bar to the
  right, mapping color to counts.  The yellow error bars are the 90\%
  confidence limits of the composite profiles centroid averaged over 5
  phase bins ($\Delta\phi=0.2$).  The red curves show the stellar
  center of mass velocities, also averaged over the same phase range
  as the data.  }
\REV{
  The composite velocity centroid very closely follows the radial
  velocity of the primary (more massive) star.  The preliminary sharp
  velocity transition at $\phi=0.5$ seen in neon
  \citep{Huenemoerder:Hunacek:2005} is not apparent in the \meg\
  curve, but the background \heg\ image does have a hint of a
  transition sharper than that due to the stellar radial orbital
  velocities.

  There are significant systematic deviations from the orbital
  velocities apparent near $\phi=0.8$.  Given the statistical
  uncertainties shown are 90\% ($1.6\sigma$), the deviations from the
  expected velocity are about $2.5\sigma$ for each grating.

  Note that the instrumental resolution is approximately $500\,
  \mathrm{km\,s\mone}$ FWHM at 12\AA\ for the \meg, and about half
  that for the \heg.  We are able to determine centroids to much
  higher precision ($\sim 25\,\mathrm{km\,s\mone}$ for \meg; $\sim
  45\,\mathrm{km\,s\mone}$ for \heg) because have accumulated signal
  over multiple profiles, and because each profile is over-sampled by
  the spacecraft dither which randomizes line placement with respect
  to pixel boundaries.  \citet{Hoogerwerf:al:2004} achieved precision
  of about $\sim 15\,\mathrm{km\,s\mone}$ from \meg\ spectra, and also
  quote a statistical verification of the precision of a Gaussian
  fitted centroid.  Also for comparison,
  \citet{Ishibashi:al:2006}\footnote{Preprint:
    http://arxiv.org/abs/astro-ph/0605383} using different techniques
  on \hetgs\ spectra of Capella, a low-orbital velocity sytem,
  obtained $3\sigma$ velocity uncertainties of about
  $20\,\mathrm{km\,s\mone}$ which is about $11\,\mathrm{km\,s\mone}$
  for 90\% confidence. It is not surprising that our precision is
  somewhat less given the lower signal per phase bin and given more
  velocity smearing in phase for this short period system.
}

\REV{
  The implications of the velocity centroid variations with regard to
  geometric distribution of the emitting plasma will be discussed in
  Section~\ref{sec:corgeom} in conjunction with light curve
  variability, emission measure, and density determinations.
}

\subsection{Line Fluxes}\label{sec:lineflux}

Line fluxes, which are required for differential emission measure (\dem)
analysis, were measured by fitting a the sum of a continuum model and
a number of Gaussians to narrow regions of the counts spectra.  The
parameters of the fits were the Gaussian centroids and fluxes.  The
continuum and Gaussian widths were determined {\it a priori}.  For the
continuum, we used a plasma model determined iteratively from the
emission measure solution (see Section~\ref{sec:dem}).  For the first
iteration, we fit a two-temperature plasma model to relatively
line-free regions, as determined by the model and instrument response.
Since lines were measurably broader than the instrumental resolution,
we determined the amount of excess line width required at one
wavelength for a strong, relatively isolated line (\ion{Mg}{12}
8.4\AA), then scaled the width with wavelength, under the assumption
that the broadening is due to either rotational or orbital effects.
However, we simply treated the broadening as a Gaussian, which was
adequate for the purposes of obtaining good fits to line fluxes and
positions.  The excess required was equivalent to about $200\,\kms$ of
turbulent broadening.

All fits were done by convolving the model with the instrument
response, using effective areas and line-responses made with \ciao\ 
tools ({\tt mkgarf} and {\tt mkgrmf}; see Section~\ref{sec:hires}).
Positive and negative orders were combined dynamically (that is, in
memory, with no new counts or response files created on disk), and the
\heg\ and \meg\ spectra were fit jointly in regions were there were
sufficient counts in each.

Line identification and blending were determined iteratively with the
\dem\ solution.  Given the plasma model, blended components were added
and removed from the fits according to whether the model showed that
resolved blends were important or negligible.

The measurement methods used here are similar to those described by
\citet{Huenemoerder:Canizares:al:2003}.  All the line fitting was done
with scripts programmed in ISIS version 1.2.8.

\begin{figure*}
%  \centering\leavevmode
%  \includegraphics[scale=1.0]{Fullspec_1.ps}
%  \includegraphics[scale=1.0]{Fullspec_1.ps}
  \epsscale{0.8}
\epsscale{1.75} % mod for preprint
  \plotone{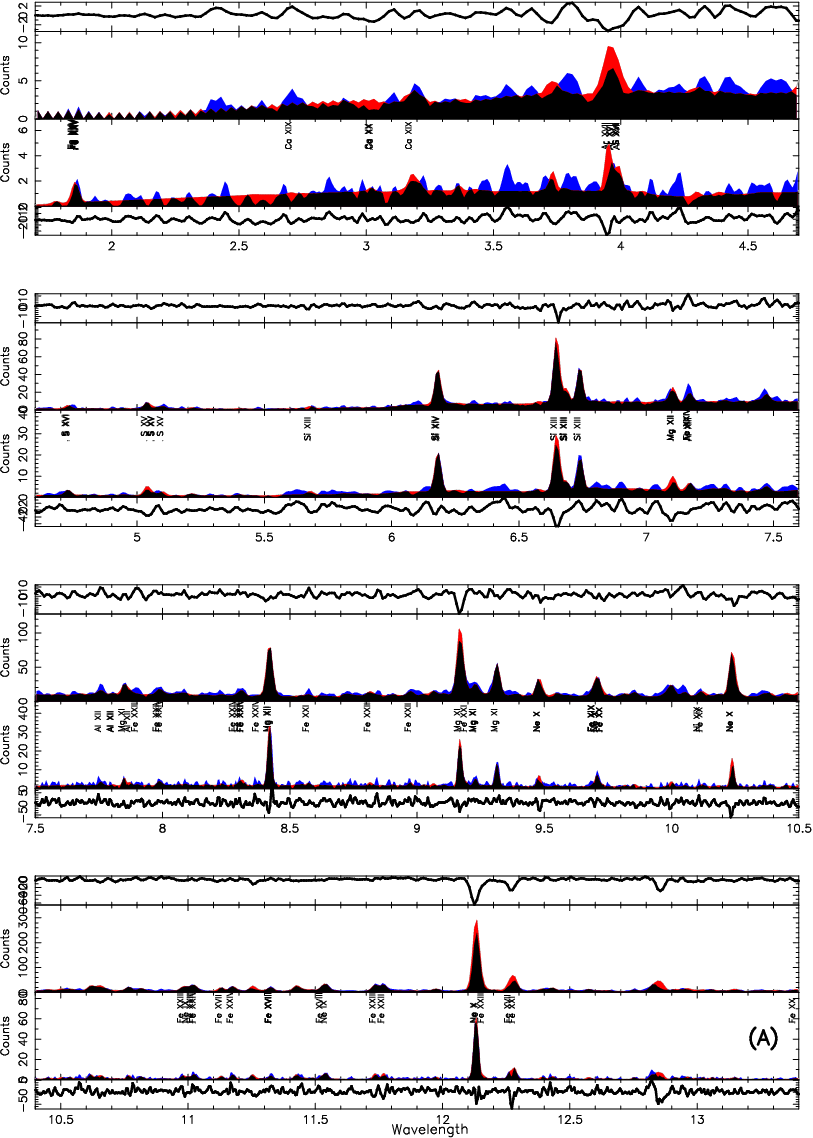}
  \caption{Detailed spectrum and synthetic spectrum from the \dem\ and
    abundance fit.  The inner two panels of each box show the \meg\  
    (upper) and \heg\ spectra.  The counts are shown in dark color
    (blue), the folded model in lighter color (red).  The overlap is
    filled with black.  The very top and bottom panels of each box are
    counts  residuals for each spectrum.  Lines from
    Table~\ref{tbl:linelist} are     marked.  Spectra were binned by
    two and Gaussian smoothed, except     for the 1.7--4.7 \AA\ region
    where both \heg\ and \meg\ were 
    grouped by 4 bins, the 4.6--7.6 \AA\ and $>13$ \AA\ regions where
    \heg\ was grouped by 4 bins.
    \label{fig:fullspec}
  }
\end{figure*}

The resulting counts spectrum and convolved model are shown in detail
in Figure~\ref{fig:fullspec}.  

\placefigure{fig:fullspec} 
%\clearpage
%\input{figfullspec}
\notetoeditor{figure~\ref{fig:fullspec} should be large - approx full
  page} 

\begin{figure*}
%  \centering\leavevmode
%  \includegraphics[scale=1.0]{Fullspec_2.ps}
\epsscale{1.75} % mod for preprint
  \plotone{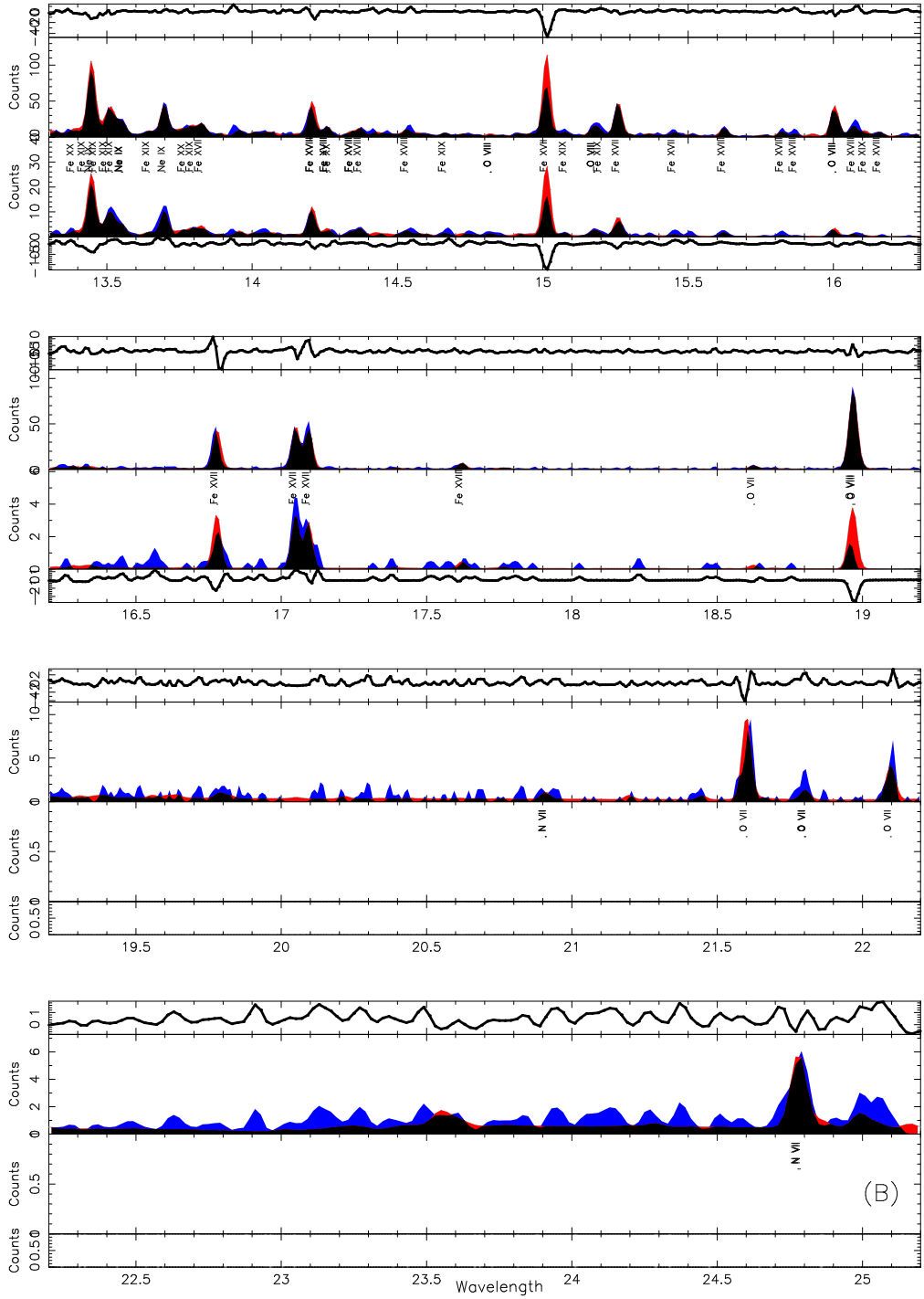}
  \caption{Detailed spectrum, part B;  see Figure~\ref{fig:fullspec}
    \label{fig:fullspecb}
  } 
\end{figure*}
\clearpage
\placetable{tbl:linelist} 
All the line measurements are given in Table~\ref{tbl:linelist}.  We
list more features than were used in analysis.  Some were rejected
because they had large wavelength residuals and so are misidentified
or are strongly blended.  Some were rejected because they had large
flux residuals and so are misidentified or have poor atomic data.  For
example, the \eli{Fe}{17} 15 \AA\ line is nearly always
under-predicted, which may be due to deficiencies in the atomic
data \citep{Laming:00,Doron:Behar:02,Gu:02}. Some lines were fit simply
to provide a good determination of the flux and wavelength of a nearby
``interesting'' feature.  The ``Use'' column of
Table~\ref{tbl:linelist} indicates which lines were used in emission
measure reconstruction, or for composite line profiles.  The density
sensitive He-like triplet inter-combination and forbidden lines were
not used in \dem\ reconstruction.

\subsection{Emission Measure}\label{sec:dem}

The differential emission measure is a one dimensional
characterization of a plasma, and can be defined as $N_e^2\,dV/dT$, in
which $N_e$ is the electron density, $V$ is the emitting volume, and
$T$ the temperature.  The $DEM$ is an important quantity because it
represents the radiative loss portion of the underlying heating
mechanism.  An emission measure can be derived from measurements of
line fluxes and some assumptions about the homogeneity and ionization
balance of the emitting plasma.  Derivation of the emission measure
relies on detailed knowledge of fundamental atomic parameters.  Even
given accurate emissivities, the contribution functions versus
temperature are broad so the emission integral cannot be formally
inverted.  Hence there are many methods to regularize the solution to
obtain the emission measure and elemental abundances.

\citet{Huenemoerder:Hunacek:2005} used a simple method described by
\cite{Pottasch:1963}, in which the $DEM$ is approximated by a ratio of
line luminosity to average line emissivity at the temperature of the
maximum emissivity.  Here we improve upon that by simultaneously
fitting the \dem\ and abundances using a method similar to that
described by \citet{Huenemoerder:Canizares:al:2003}, who also
discussed some of the caveats of emission measure reconstruction and
gave relevant citations.  Briefly, the relation between emissivity and
flux is an ill-posed problem.  To obtain a unique solution, under
assumptions of the model, a regularization term of some form must be
included.  Here we have replaced explicit smoothing of the \dem\ in
our prior work with a regularization term, so that the
functional form of the statistic we minimize is
\begin{eqnarray}
   \label{eq:demchisq}
   \chi^2&=\sum_{l}
             \frac{1}{\sigma_l^2}
             \left[L_l - A_{Z(l)}
               \sum_{t}{\delta_t\epsilon_{lt} D_t} \right]^2 %\\
             + q P(D),% \nonumber
\end{eqnarray}
in which $l$ is a spectral feature index and $t$ is the temperature
index.  The measured quantities are the line luminosities, $L_l$, with
uncertainties $\sigma_l$. The {\em a priori} given information are the
emissivities, $\epsilon_{lt}$ as defined by \citet{Raymond:96},
$\delta_t$ is the logarithmic bin size, and the source distance (which
is implicit in $L$).  The minimization provides a solution for the
differential emission measure, $D$, and abundances of elements $Z$,
$A_Z$.  To naturally constrain the \dem\ to be positive, we actually
fit $\ln D_t$.  $P(D)$ is a regularization term (or ``penalty
function'') which is only a function of the model, and $q$ is a scale
factor which specifies the relative importance of the regularization.
For $P$, we used the sum-squared second derivative of the \dem\ with
respect to $\log T$, and a multiplier to make the two terms of the
statistic comparable.  This form imposes a minimum smoothness on the
solution; the \dem\ cannot have large changes in curvature.  We did
not impose any regularization on the abundances, since we have no
intuitive bias on their functional form.

The \dem\ fitting was done with custom software in ISIS. While ISIS
has no built-in \dem\ reconstruction, it does have sophisticated
plasma database access and evaluation functions permitting efficient
lookup and  computation of emissivities.  It also provides for
user-defined fit functions and statistics, which made it
straightforward to develop a \dem\ model within its fitting and
modeling infrastructure.

As an initial guess, we assumed cosmic abundances and a boxcar \dem;
the amplitude of the central temperature region was chosen to
approximately produce the observed line counts, and the ends were set
to a very low value.  The reason for leaving the \dem\ at the
temperature extrema low is that the \dem\ is not well constrained
\REV{
  physically there by lines, but the hot portion does strongly affect
  the continuum.  Hence, we preferred a solution in which emission
  weights are increased only if required by the fit to lines alone.
  The normalizations outside our regime of temperature sensitivity,
  roughly $\log T = 6.4$ (\eli{O}{7}) to $7.8$ (\eli{Fe}{25}; the tail
  of \eli{Si}{16}), were artificially constrained to have negligible
  emissivity.
}
The line-to-continuum ratio was fit {\it post facto} by scaling the
normalization of the \dem\ and abundances.

The model \dem\ was obtained iteratively.  For each trial \dem, we
generated a new continuum model and re-fit the lines.  After the second
such iteration, we reviewed line residuals and excluded those which
were very large (greater than $3\sigma$), under the assumption that
they were misidentified or blended.  The final fit included only the
selected lines, and was {\it post facto} renormalized to reproduce the
observed line-to-continuum ratios in the 8-11 \AA\ range.  The line
fluxes predicted by the \dem\ and abundance model are listed in
Table~\ref{tbl:linelist}, along with the residuals from the parametric
fit, and their $\delta\chi$ value (last three columns).  
\REV{
The ``Use'' column flags lines used in the \dem\ reconstruction with
an ``E''.
}
Due to the iterative nature of the model, the measured line fluxes are
slightly dependent upon the \dem\ solution through the definition of
the continuum.

\begin{figure}
%  \plotone{DEM.ps}
\epsscale{1.0} % mod for preprint
  \plotone{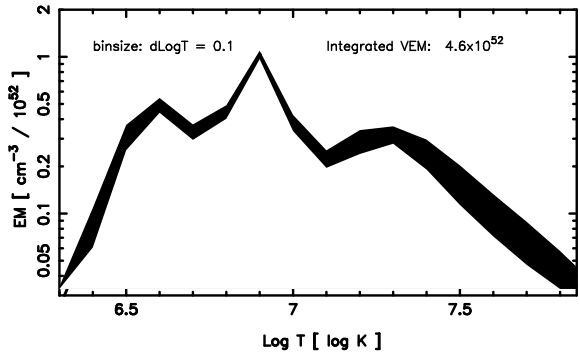}
  \caption{In this reconstructed emission measure distribution, the
    width of the filled region is the 1-standard-deviation range as
    determined from a Monte-Carlo run in which the observed line
    fluxes were randomized according to their statistical uncertainty.
\REV{The fit is physically unconstrained at the highest and
      lowest temperatures plotted, but was artificially constrained to
      have a negligible value at the extrema.
}
The integrated emission measure is about
$4.6\times10^{52}\,\mathrm{cm\mthree}$. 
    \label{fig:dem}
  }
\end{figure}

\placefigure{fig:dem}    %\input{figdem}
%\clearpage
%\clearpage
%
%
Figure~\ref{fig:dem} shows our final emission measure (integrated over
logarithmic bins of width 0.1 dex), using the Astrophysical Plasma
Emission Database (\aped, \REV{Version 1.3.1}) for line emissivities
\citep{Smith:01}, the ionization balance of \cite{Mazzotta:98}, and
Solar abundances from \cite{Anders:89}.  The envelope shown was
determined by computing the deviation in solutions over a Monte-Carlo
run of about 100 fits in which the line fluxes were varied according
to their measured statistical uncertainties, assuming a Gaussian
distribution about their fitted values.  This assumption would
actually result in an enlarged uncertainty bounds, since the data are
an estimate of the true mean, and the random perturbation would
sometimes be toward the true mean, and sometimes away, perhaps further
than the the measured uncertainty and truth would allow.  

\REV{
  Other sources of systematic error, the uncertainties in the atomic
  physics and in the instrumental calibration, were not included, and
  probably accounts for a similarly sized envelope.  While systematic
  uncertainties in atomic data are difficult to incorporate
  explicitly, \citet{Huenemoerder:Canizares:al:2003} applied an
  approximate method, in which they set a global lower limit on the
  line flux uncertainty of 25\% (regardless of the statistical
  uncertainty) and repeated the \dem\ reconstruction.  They found a
  similar shaped distribution, but proportionally larger uncertainty.
  The important quantity for understanding coronal activity is the
  overall shape of the \dem, which seems reliable under the current
  assumptions and calibration uncertainties.
}

There is sharp structure in the \dem\ with a large peak at $\log
T=6.9$, a second peak at 6.6, and hot tail from about 7.2 to 7.8.
This is qualitatively similar to the simple provisional \dem\ of
\citet{Huenemoerder:Hunacek:2005}, but with sharper structure as
expected from an iterative solution.

\begin{figure}
%  \plotone{Abund.ps}
\epsscale{1.0} % mod for preprint
  \plotone{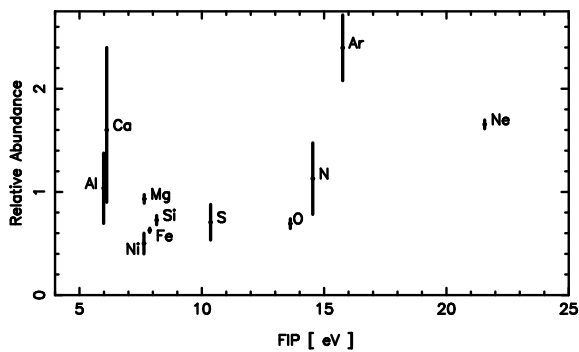}
  \caption{The reconstructed abundances relative to Solar are plotted
    against their first ionization potential (FIP).  Error bars are
    the one standard deviation range from the Monte-Carlo emission
    measure and abundance reconstruction, or in the cases of Ni and
    Ca, from a {\it post facto} fit of abundances given the $DEM$.}
    \label{fig:abund}
\end{figure}

\placefigure{fig:abund}  %\input{figabund}
%
%\clearpage
%\clearpage 
The fitted coronal abundances, relative to Solar (since we
do not know the stellar photospheric values), are shown in
Figure~\ref{fig:abund}.  The uncertainties are the $1\sigma$ range as
determined by the Monte-Carlo iteration, except for Ni and Ca.  The
latter two were not included in the $DEM$ solution since they were
weak and were not fit with parametric profiles as were the strong
lines (and so they do not appear in Table~\ref{tbl:linelist}).
Instead, we fit them {\it post facto} by using the $DEM$ and abundance
solution to define a plasma model, then adjusted the abundance of Ca
\REV{by minimizing the residuals between the binned counts and
  synthetic spectrum in the \ion{Ca}{19} He-like triplet near 3.2 \AA\
  as a function of Ca abundance.  }
There is a blend of \ion{Ar}{17} here, but given the $DEM$, it is
expected to be about an order of magnitude weaker than Ca.  For Ni, we
similarly fit the \ion{Ni}{19} lines at 14.043 and 14.077 \AA, and
\ion{Ni}{19} 12.435 \AA.  In each region, blended Fe lines had model
fluxes 5-10 times weaker than the nickel lines. 
\REV{
The Ca abundance uncertainty is large because there are few counts and
a relatively strong continuum.  The Ni has smaller uncertainty because
there are more counts and weaker continuum.
}

\subsection{Density}

The helium-like triplets are well known as density sensitive
diagnostics, particularly the ratio of the forbidden ($f$) to
inter-combination ($i$) line fluxes, commonly known as the $R$-ratio,
$R = f/i$ \citep{Gabriel:69,Porquet:00,Porquet:01,Ness:Mewe:al::2001}.
A mildly temperature sensitive diagnostic is the $G$-ratio, which is
the sum, $f+i$, divided by the resonance line ($r$) flux.  The
critical density increases with atomic number.  In the \hetgs\ range,
the ions of interest for coronal diagnostics are \ion{O}{7},
\ion{Ne}{9}, and \ion{Mg}{11}.

The \ion{O}{7} triplet ratios are relatively straightforward to
determine since the lines are well separated, unblended, and the
continuum is low.  \ion{Ne}{9} is a difficult case since it is blended
with several line of \ion{Fe}{19} and \ion{Fe}{20}, in addition to
having a significant continuum.  \citet{Ness:Brickhouse:al:03}
performed a thorough analysis of this region in Capella at different
resolutions, and showed that \heg\ resolution is necessary for
accurate modeling.  The \ion{Mg}{11} region is of intermediate
difficulty; it contains blends with the \ion{Ne}{10} Lyman-like
series, which converges near the inter-combination line.  Neon is often
overabundant in stellar coronae, making the contribution possibly
significant.

\begin{figure}
%%  \plotone{cO7GR_sp_v2.ps}
  \epsscale{0.4}
%  \plotone{cO7GR_sp_v2-letgs.ps}
\epsscale{1.0} % mod for preprint
  \plotone{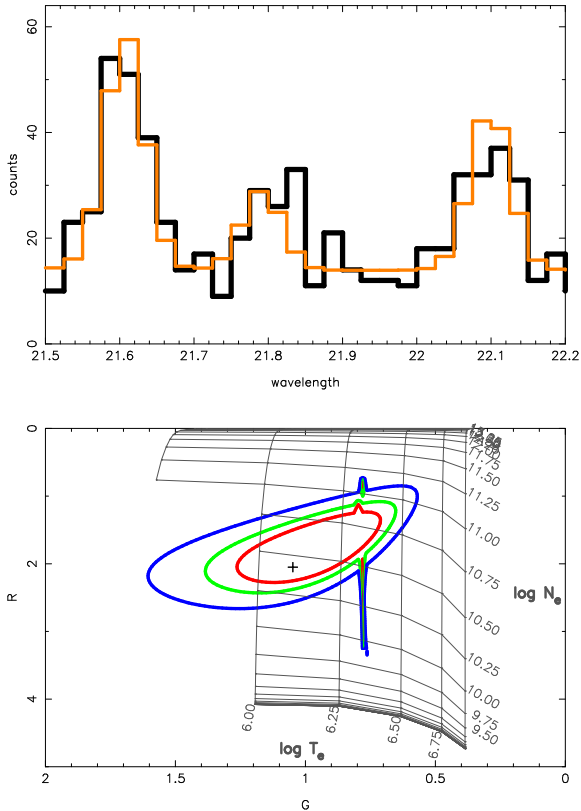}
%  \centering\leavevmode
%  \includegraphics[scale=0.75]{cO7GR_sp_v2-letgs.ps}
  \caption{
    The \letgs\ \ion{O}{7} spectral region (top, dark histogram),
    model (top, light histogram), and the confidence contours (bottom,
    colored ovals) for the $G$ and $R$ ratios.  The theoretical
    isothermal, isodensity $G-R$ ratio grid is plotted in gray, and
    labeled with the logarithmic density and temperature values.  The
    best fit density is about $\log N_e = 10.5$.  The \hetgs\ data
    give a similar result, but with larger contour regions.  (The
    vertical spike in the contours is a numerical artifact.)}
    \label{fig:ogrcontours}
\end{figure}

We have determined $R$ and $G$ ratio confidence contours for
\ion{O}{7} and \ion{Ne}{9} by directly fitting the line ratios to the
spectra.  Fits were done similarly to the lines (see
Section~\ref{sec:lineflux}), but with the addition of parameter
functions to define fit-parameters for the ratios themselves.
\placefigure{fig:ogrcontours}
%\clearpage
%\clearpage
Figure~\ref{fig:ogrcontours} shows the result for oxygen. Here we have
decided to use the \letgs\ data (observation ID 2559), since it has
better signal at \eli{O}{7}, and the contours are smaller than for the
\hetgs.  Though at a different epoch, the solutions from \hetgs\ and
\letgs\ are equivalent.  We have plotted the axes reversed, since the
temperature and density increase inversely with the ratios.  The
contours show the one-, two-, and three-sigma confidence intervals,
with the best fit marked with a ``plus'' sign.  The underlying grid
shows the theoretical ratios' lines of constant temperature and
density (density dependent tables are from Brickhouse, private
communication).  We also show the counts spectrum and best fit model.
The $R$ ratio contours clearly show a density above the low-density
limit; the best fit value is $\log N_e=10.5$, with the
\REV{
  $1\sigma$ confidence interval ranges from 10.3 to 10.8.  The
  $G$-ratio spans a broad range but is not inconsistent with the
  reconstructed \dem.
}

\begin{figure}
  \epsscale{0.4}
\epsscale{1.0} % mod for preprint
%  \plotone{Ne9_reg_fit.ps}
  \plotone{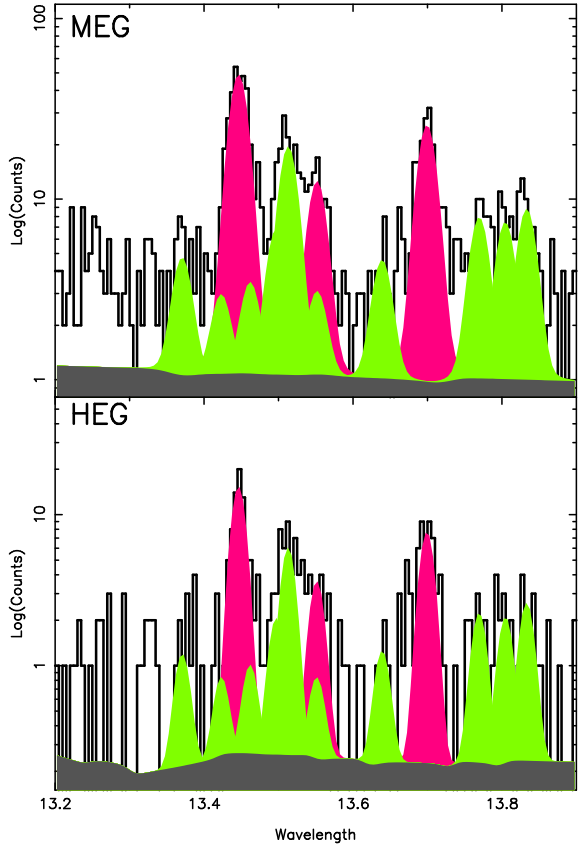}
%  \centering\leavevmode
%  \includegraphics[scale=0.75]{Ne9_reg_fit.ps}
  \caption{\ion{Ne}{9} region for \meg\ (top) and \heg\ (bottom),
    showing the counts (dark histogram), the \ion{Ne}{9} components
    (red-filled Gaussians), the iron components (green-filled
    Gaussians), and the continuum model (gray-filled region).  The
    relative intensity of the iron line blended with the \ion{Ne}{9}
    inter-combination line was determined from neighboring iron lines.
    Note that the counts scale is logarithmic.}
    \label{fig:netrip}
\end{figure}

For the \ion{Ne}{9} region, we included nine resolved lines of
\ion{Fe}{17}, \ion{Fe}{19}, and \ion{Fe}{20}.  There is
an unresolved blend of \ion{Fe}{19} and \ion{Fe}{20} with the
inter-combination line.  From fits to the neighboring lines, the \dem\ 
model and \aped, we can estimate the contribution of iron to the Ne
$i$-line of about 15\%.  This was included as a frozen component in
the ratio fit.  We show the counts, \ion{Ne}{9}, and Fe model in
Figure~\ref{fig:netrip} for both \heg\ and \meg, since resolution is
critical. The figure also shows the model continuum, and
Table~\ref{tbl:linelist} flags the resolved iron blends included.
\placefigure{fig:netrip} 
%\clearpage
%\clearpage
%
%
For the unresolved blends, we give the identifying information from
the \aped, which labels lines uniquely by the element, ion,
upper ($u$), and lower ($l$) levels: \ion{Fe}{19} $\lambda13.551$
($u=65$, $l=1$); \ion{Fe}{20} $\lambda13.535$ ($u=109,107$, $l=7$),
$\lambda13.558,13.565$ ($u=110,109$, $l=8$).

\begin{figure}
  \epsscale{0.4}
%  \plotone{cNe9GR_sp_v2.ps}
\epsscale{1.0} % mod for preprint
  \plotone{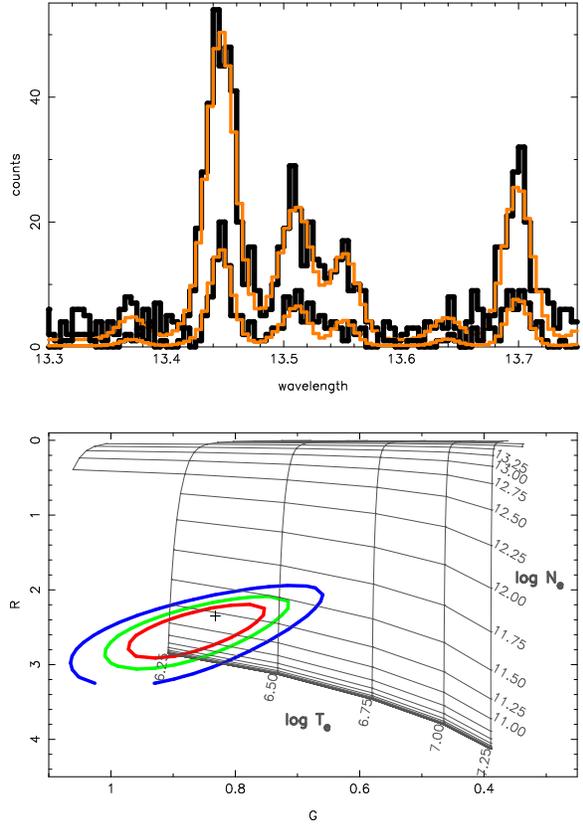}
%  \centering\leavevmode
%  \includegraphics[scale=0.75]{cNe9GR_sp_v2.ps}
  \caption{
    \ion{Ne}{9} spectral region (top, dark histogram), model (top,
    light histogram), and the confidence contours (bottom, colored
    ovals)  for the $G$ and $R$ ratios. The theoretical isothermal,
    isodensity $G-R$ ratio grid is plotted in gray, and labeled with
    the logarithmic density and temperature values.  The best-fit
    density is about $\log N_e = 11.25$.
}
    \label{fig:negrcontours}
\end{figure}

In Figure~\ref{fig:negrcontours}, we show the resulting $G-R$ contours
and spectra with best-fit model. Here, it is not clear that the
density is above the critical value; the best fit is $\log N_e=11.25$,
and the one-sigma range is 10.0-11.4.  
\REV{
  The $G$-ratio contours extend to low temperatures, but the range is
  not inconsistent with the \dem.
}
\placefigure{fig:negrcontours}
%
%\input{fignegr}
%
%\clearpage
%
%\input{figmgtrip} 
%
%\clearpage
\begin{figure}[h]
  \epsscale{1.0}
%  \plotone{vwcepMg.ps}
  \plotone{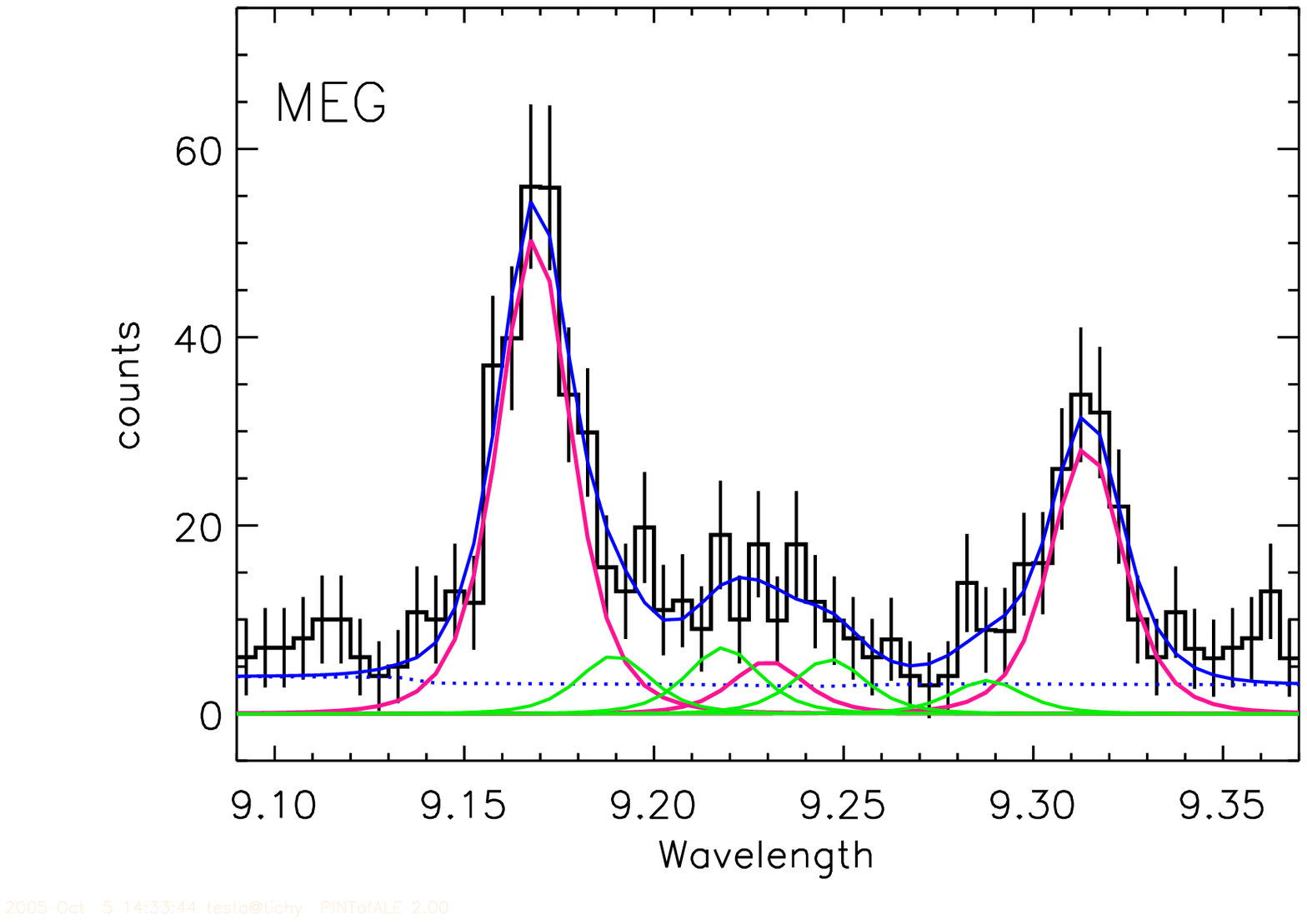}
  \caption{\ion{Mg}{11} region for the \meg,
    showing the counts (dark histogram), And individual ionic
    components as smooth curves: \ion{Mg}{11} (red), \ion{Ne}{10} and
    \ion{Fe}{00} (green)}
    \label{fig:mgtrip}
\end{figure}
%\clearpage
We also fit the \ion{Mg}{11} ratios.  The ratio and uncertainty gives
$R = 2.7 \pm 0.5$.  The theoretical low-density ratio is 2.9.  Hence,
we are well below the limit of density sensitivity (the critical
density is about $\log N_e = 13$).  We included the high-$n$
\ion{Ne}{10} Lyman-like series and iron blends in the triplet region
\citep{Testa:Drake:al:2004b}, which we show in Figure~\ref{fig:mgtrip}.
\placefigure{fig:mgtrip}

%%%%%%%%%%%%%%%%%%%%%%%%%%%%%%%%%%%%%%%%%%%%%%%%%%%%%%%%%%%%%%%%%%%%%%%

\section{Discussion}

There is no question that coronal activity saturates at short periods,
and that super-saturation occurs at even shorter periods.  Why
saturation occurs is still an open question among hypotheses of dynamo
suppression by tidal interactions, coronal stripping via Coriolis
forces, filling factors near unity, or other effects.  With high
resolution X-ray spectra, we cannot yet solve this problem, but we can
provide additional details from individual case studies.  We can make
several deduction from the VW~Cep light curves, line velocities,
emission measures, and density diagnostics.  We will also consider the
similar super-saturated W~UMa system, 44~Boo.\footnote{44 Boo is also
  officially designated ``i Boo'', but sometimes mistakenly called as
  ``44i Boo.''}

\subsection{Coronal Geometry }\label{sec:corgeom}

From visual inspection of the \vwc\ X-ray light curve, we see that
X-rays vary in brightness quasi-sinusoidally at the period of the
binary with a modulation of about 20\%.  The shape is very different
from the optical curve, which is determined primarily by the geometric
configuration of two stars with nearly equal surface brightnesses
(tidally distorted spheroids which undergo partial eclipses) and
secondarily by inhomogeneous brightness distributions (gravity
darkening and starspots).  Since we do not see obvious eclipses in
X-rays (which could have a phase duration of about 0.2), we infer that
the regions obscured probably do not contribute significant X-ray
emission within our detection sensitivity.  Since the inclination is
$63^\circ$, and the primary has about twice the radius of the
secondary, the regions covered are roughly the southern hemispheres of
each star.  The north-polar region of each star is always visible,
roughly within co-latitude of $30^\circ$, so any emission from this
region on either star will contribute a constant emission component
(within projected area affects).  The non-uniform emitting material
can thus be roughly constrained to latitudes of $0-60^\circ$, in one
hemisphere only, with more emission measure seen near phase 0 (larger
star in front).

If we now consider the X-ray composite radial velocity curve, we can
rule out significant emission from the secondary (smaller) star, at
the limit of about 30\%.  The ratio of photospheric surface areas is
about 2:1.  If we had 30\% of the X-ray line flux originating from the
secondary, it would have been apparent in the composite line profiles,
but we do not see it (see Figure~\ref{fig:clp}).  We explored the
limits of the secondary's signature in the composite profiles through
simulations.  We simulated spectra using the same model emission
measure for each star, but doing a weighted sum at each phase, with
each spectrum redshifted according to the phase.  We used the same
exposure times from each bin of the observed phase curve as well as
ones 100 times larger.  We considered relative weights, primary to
secondary of 0.7:0.3 (the ratio of areas), 0.8:0.2, and 1:0.  After
the simulated data were produced, the composite lines were formed and
measured in exactly the same way as for the observation.

For the 0.7:0.3 weighting, modulation in the composite centroid is
reduced to an undetectable size.  For 0.8:0.2, modulation is reduced
by about 50\%, to about the maximum deviation seen in the \heg\ near
phase 0.8.  For all emission from the primary, the curve follows the
primary radial velocity, as expected.  Thus, we find most emission at
most phases can be attributed to the primary.  At some phases,
approximately 0.75--1.0, we may have about 20\% from the secondary.
Since there is no apparent eclipse of the secondary (at phase 0.0, the
primary eclipse), the emission from the secondary must be at high
latitudes on the trailing hemisphere, but not polar or we would have
a larger velocity centroid perturbation near phase 0.25.

We can next apply emission measure and density information to estimate
coronal extent, under some simplifying assumptions about the emitting
plasma's geometry.  If we assume that the 20\% light curve intensity
modulation is caused by a random distribution of identical
\REV{coronal structures} of constant cross-section, then we can have
about 25 \REV{structures}, based purely on counting statistics.  Using
the scaling relation of \citet{Huenemoerder:01}, an integrated
emission measure of $4.6\times10^{52}\,\mathrm{cm}\mthree$ and a
density range of $\log N_e = 10.5$ to $11.25$, and assuming all
emission comes from the primary of radius $0.9\,R_\odot$, then the
\REV{structure} height is about 0.06 to 0.2 stellar radii.  Such a
corona would thus be fairly compact, which is self consistent; if it
were very extended, flux modulation could be much less.  A single such
\REV{structure} could have a height from 0.2 to 0.6 stellar radii and
could probably be placed to create the same intensity modulation.  We
prefer the multiple \REV{structure} hypothesis, but this is by no
means a unique requirement of the data.  If we spread the emitting
volume more uniformly over a hemisphere, then for the above density
range the fractional height is 0.02 to 0.03.  We probably have a
compact, polar corona.

\citet{Gondoin:2004a} reported an X-ray eclipse seen \xmm\ light curves
obtained 10 months prior to the \cxo\ observations and concluded that
both components were X-ray emitters.  If the dip in \xmm\ rate were
really an eclipse, then this would indicate rather large and rapid
changes in coronal structure.  The \xmm\ observation, however, did not
cover a single continuous orbit, and the dip is offset from phase 0.0,
so the interpretation is not definitive.  Also contrary to our
results, \citet{Gondoin:2004a} derived loops quite large in comparison
to the stellar radii (20-80\%), but admittedly by methods which are
not self-consistent.

\subsection{Temperature Distribution, Heating, Opacity }

The differential emission measure (Figure~\ref{fig:dem}) is similar to
other active stars, being highly structured and spanning a broad range
in temperature.  \citet{Sanz-Forcada;Brickhouse:02, SanzForcada:al:03}
show a large collection of \dem\ distribtions with a variety of peaks,
bumps, and tails.  In our analysis, we included the flare times and
this no doubt produces the bump above $\log T = 7.1$, similarly to the
results for II~Peg \citep{Huenemoerder:01} and Proxima Cen
\citep{Gudel:Audard:al:2004}, which were modeled at different activity
levels. 

The most prominent peak in the \vwc\ \dem\ is well defined and narrow,
with a slope approximately proportional to $T^4$.  Such a feature has
also been observed in other stars of different types
\citep{Brickhouse:00,Sanz-Forcada;Brickhouse:02}.  The \dem\ can give
us some insights into the underlying physics.  A possible
interpretation of this recurrent feature is linked to the spatial
location and temporal distribution of coronal heating.
\citet{Testa:al:2005} have proposed a hydrodynamic model of loops
undergoing pulsed heating at their foot-points, which is able to
reproduce the presence of a peak and the steep rise of the \dem\
observed in these active stars.  Whatever the mechanism, the heating
in VW~Cep does not look radically different from other coronal
sources, at least as manifested in the \dem.

The abundances do not show any strong trend with the first ionization
potential (FIP).  While Ar looks unusually abundant in
Figure~\ref{fig:abund}, the statistical uncertainty is large.
Furthermore, the measurement is also subject to systematic uncertainty
in placement of the continuum.  The Ar abundance is probably not
significantly different from Ne.  The overall distribution looks very
similar, for example, to that of the RS~CVn binary, AR~Lac
\citep{Huenemoerder:al:2003}.  The Ne:O ratio which we derived from
the iterative emission measure and abundance reconstruction is $0.35
\pm 0.03$. This is comparable to the rather robust active star mean of
$0.41$ found by \citet{Drake:Testa:2005} from temperature-insensitive
line ratios.  From the abundance analysis, we again have no reason to
think that the corona of a W~UMa system, though super-saturated,
differs from other coronally active stars.

We examined line ratios which have been used for opacity diagnostics.
\eli{Fe}{17} has been problematic, because the ratios systematically
differ from theoretical values \citep{Laming:00,Doron:Behar:02,Gu:02}.
\vwc\ is no exception; its \eli{Fe}{17} 15.01 \AA\ line is
over-predicted.  We find no significant differences in ratios from
typical optically thin values shown in \citet{Testa:Drake:al:2004a} or
\citet{Ness:Schmitt:al:2003}.

\subsection{ Comparison to 44 Boo }

\ffb\ is a contact binary system which is very similar to \vwc, having
similar period, mass ratio, and spectral types.  Its fundamental
characteristics are given by \citet{Hill:Fisher:al:1989} and a
velocity curve by \citet{Lu:Rucinski:al:2001}.  \citet{Brickhouse:01}
have analyzed \hetgs\ spectra and concluded that most emission is at
high latitudes, compact, that there is a very compact emitting region
on the smaller star (the secondary) which causes a very brief dip in
X-ray brightness when it momentarily rotates out of view, and a larger
emitting region near the pole of the primary which gives rise to a
quasi-sinusoidal brightness modulation, but which isn't eclipsed.
\citet{Gondoin:2004b} presented \xmm\ observation of \ffb\ which
covered on orbital period. Based on the absence of eclipses, he
concluded that the corona must be extended, though he presented other
arguments which led to compact loops.

\begin{figure}
%    \plotone{44b_tglc.ps}
\epsscale{1.0} % mod for preprint
    \plotone{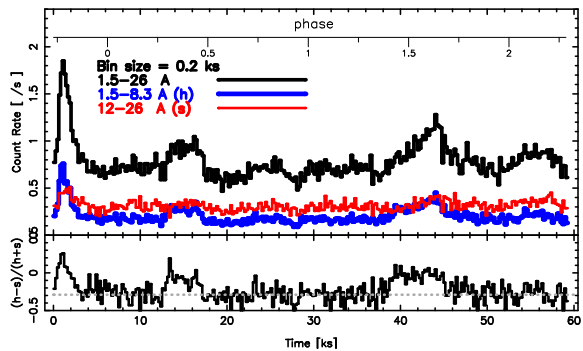}
  \caption{Count rates of 44~Boo in 200 s bins.  In the upper panel
    are light curves extracted from all diffracted photons (1.5-26\AA;
    upper heavy curve), a ``soft'' band (12-26\AA; lower, thinner
    curve), and a ``hard'' band (1.5-8.3\AA; lowest, thicker curve).
    \REV{Near the top is an axis giving the orbital phase plus a
      rotation count.}
    The lower panel shows a hardness ratio (solid histogram), and the
    median of the hardness for the middle flat section (light dotted
    line).  We defined the three high hardness regions as flares and
    excluded them from the phase curve (0-5, 12-18, and 38-46 ks).
    \label{fig:ffblc}
  }
\end{figure}

\begin{figure}
%  \plotone{44b_tgpc.ps}
\epsscale{1.0} % mod for preprint
  \plotone{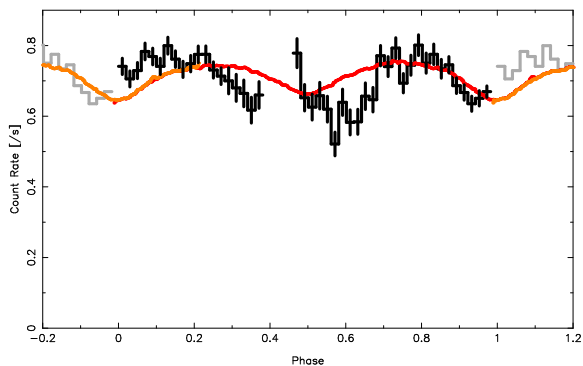}
  \caption{ The dark, solid histogram is the phase-folded X-ray
    count-rate curve of 44~Boo, excluding flare times.  Bins are
    $\Delta\phi=0.025$, or 462 s, with a cumulative exposure per bin
    of about 460 to 1400 s, depending on the phase.  The gap near
    phase 0.4 is from missing data, due to flare-time filtering.  The
    smooth curve is a scaled $B$ light curve to compare optical and
    X-ray light spatial distributions. Photometric $B$ data are from
    T.\ Pribulla ({\tt
      http://www.astro.sk/$\sim$pribulla/lc/44boo.jpg}.)
    \label{fig:ffbpc}
  }
\end{figure}

Using the same techniques applied to \vwc, we have extracted the X-ray
light and phase curves and a composite line velocity curve from the
same \hetgs\ observation used by \citet{Brickhouse:01} (Observation ID
14). The light curve in Figure~\ref{fig:ffblc} shows three flares.
The rise in the final 10 ks, identified as a flare by
\citet{Brickhouse:01}, shows no increase in hardness, so we accept it
into the phased curve.
\placefigure{fig:ffblc}  %\input{fig_44b_lc}
%\clearpage
%\clearpage
%
Our phased X-ray light curve (Figure~\ref{fig:ffbpc}) is
qualitatively different from \citet{Brickhouse:01}.
\placefigure{fig:ffbpc}   %\input{fig_44b_pc}
%
%\clearpage
%\clearpage
%

Since W~UMa stars undergo period variations, we were careful to
reference the ephemeris to light-curves from the same epoch
\citep{Pribulla:al:2001}.  We also obtained a $B$ photometric light
curve from the same epoch and scaled it to arbitrary relative
intensity to display over the X-ray curve.  We do find strong evidence
for the primary eclipse in the broad dip in X-ray flux in good
correspondence with the $B$ intensity.  The rest of the curve is
qualitatively similar to the \vwc\ X-ray curve (Figure~\ref{fig:pc}),
having a rather broad depression with no obvious eclipse. The gap
beginning near phase 0 is due missing data due to the exclusion of
flares.  We do not find a very narrow dip nor require a very small
near-polar emitting region as hypothesized by \citet{Brickhouse:01}.
These differences are probably primarily due to our forming the
histogram in phase bins, weighting the counts by the exposure function
(which ranges from zero to 2 ks), and in our flare discrimination
using a hardness ratio.

\begin{figure}
  \epsscale{0.4}
\epsscale{1.0} % mod for preprint
  \centerline{\plotone{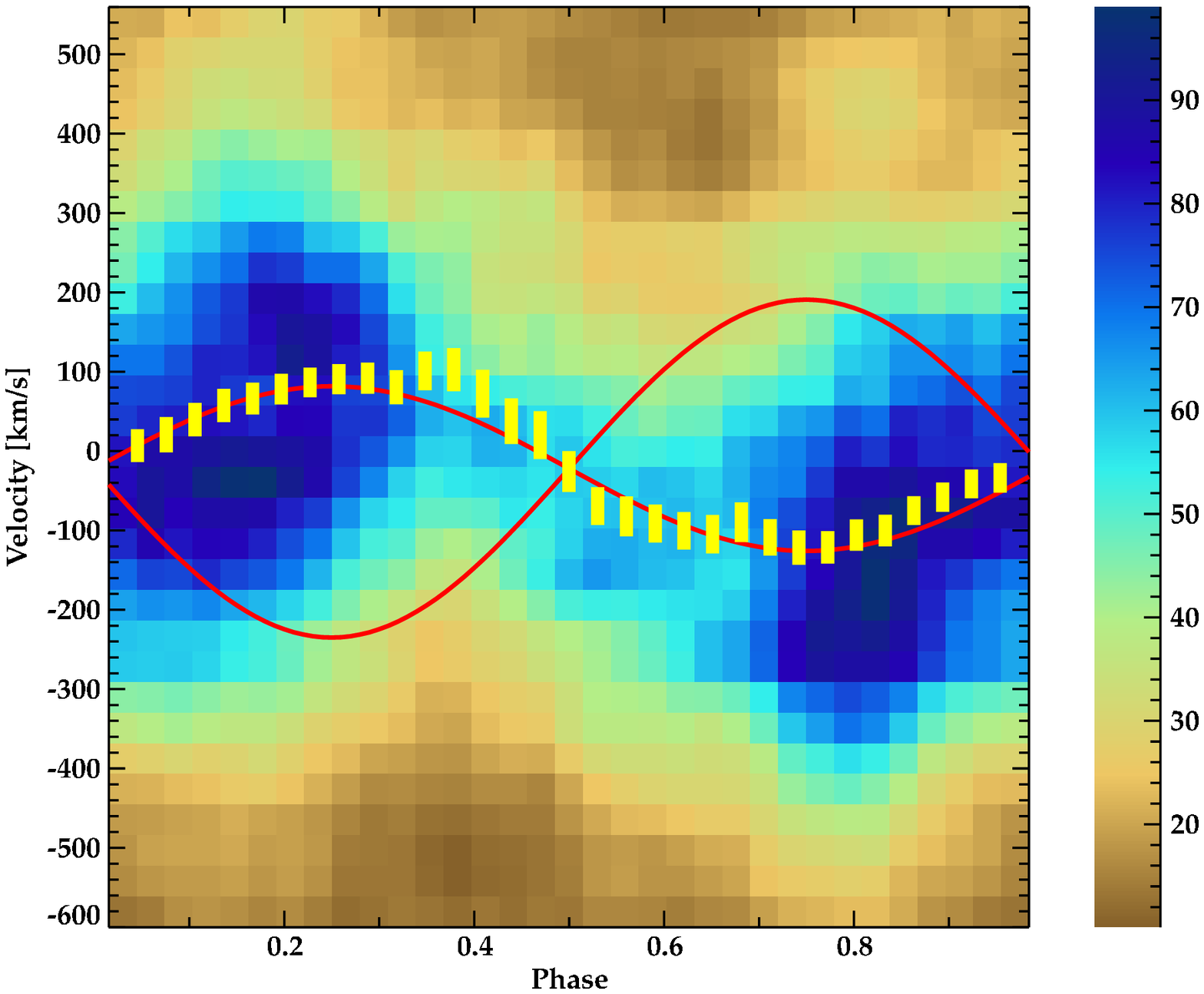}}
%  \newline
%  \includegraphics[scale=0.4]{f15b.eps}
\epsscale{1.0} % mod for preprint
  \centerline{\plotone{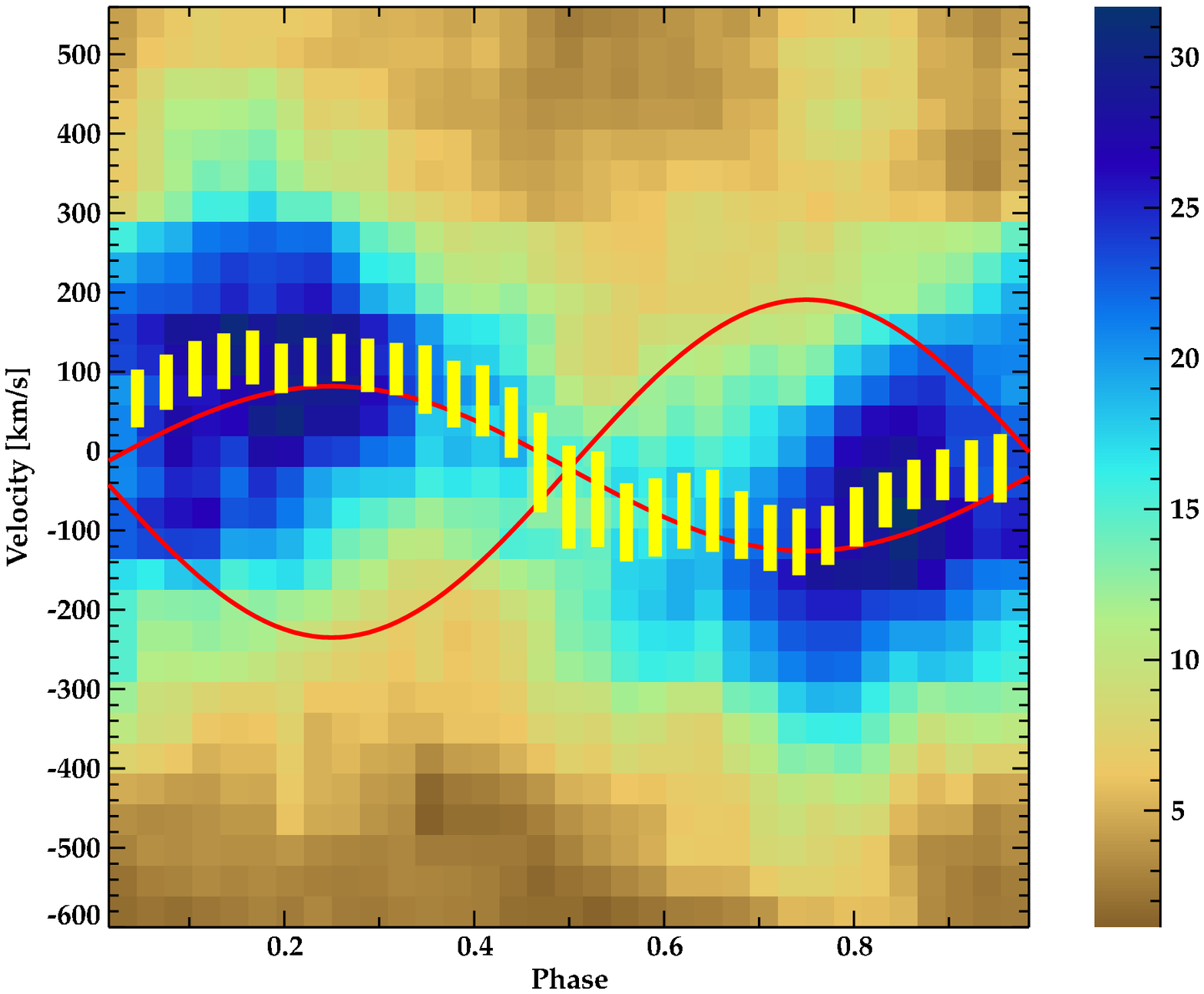}}
  \caption{
    Composite line profile for the MEG spectrum (top) and HEG (bottom)
    for 44 Boo. Yellow bars are the 90\% confidence intervals of the
    centroid of the composite profile. The red sinusoidal curves are
    the center-of-mass velocities of the binary components.  The
    background is an intensity map of the composite profiles' counts.
    The bar to the right of each image gives the counts-to-color
    mapping for the image.  The centroids are correlated since they
    were measured in phase bins of width 0.2. The red curves were
    averaged over the same intervals.
    \label{fig:ffbclp}
  }
\end{figure}

Figure~\ref{fig:ffbclp} shows our composite line profiles and
centroid, which is qualitatively similar to the velocity curve of
\citet{Brickhouse:01}.
\placefigure{fig:ffbclp}   %\input{fig_44b_clp} 
\notetoeditor{figure~\ref{fig:ffbclp} should be one column wide, w/ 2
  parts stacked vertically}
%\clearpage
%\clearpage
%
The primary clearly dominates the composite line centroid. However, the
distortions and wings also hint at some contribution from the
secondary, particularly between phases 0.6--0.9 where the centroid is
systematically shifted to more positive velocities.  The secondary
cannot contribute as much as half the light, however, or the composite
profile centroid would be strongly biased toward zero.  Given the
presence of the primary eclipse, the broadening in the profile wings,
slight distortions in the \heg\ centroid, and simulations, we estimate
that the secondary contributes no more than about 20\% of the X-ray
flux during some phases.

Were it not for the primary eclipse, the \ffb\ situation would be very
much like \vwc: quasi-sinusoidal light curve with a minimum near
phase 0.5, 20\% contribution from the secondary at some phases.  \ffb\
is at a slightly higher inclination than \vwc,  $73^\circ$ vs.\
$63^\circ$.  This does give it a higher probability of eclipsing
compact coronal emission concentrated at high latitudes.  On the other
hand, it is highly possible that the size and location of the coronal
emission migrates, as do the photospheric spots (see, for
example, \citet{Hendry:Mochnacki:2000}).  At future epochs, it is
possible that \vwc\ will show X-ray eclipses.

\section{Conclusions}

Super-saturation, at least in the cases of VW~Cep and 44~Boo, appears
to be manifested in small area of coverage by compact, near-polar
structures.  Given roughly a factor of two for coronal structures
occurring in only one hemisphere, and a factor of two in expected
emission per area (the secondary is weak relative to the primary), the
X-ray flux can be significantly depressed below the saturated limit of
$L_\mathrm{x}/L_\mathrm{bol}\sim10^{\mthree}$.  {\em Why} this occurs
is still unexplained, but we favor the general scenario of
\citep{Stepien:Schmitt:Voges:2001} which forms polar, but compact,
structures.

Given the small area of corona on the secondary, we predict that the
secondary's coronal signature be highly changeable with phase over
different epochs.  This would show as changes in the depth of primary
eclipse and phase of perturbation in the velocity profile.  The
\hetgs\ is the only X-ray spectrograph currently capable of making
these measurements.  Obtaining high signal-to-noise per phase bin is
of utmost importance to surface geometry reconstruction techniques.
Further multi-orbit spectroscopy of these or other short period
systems is certainly important to determination of transient or common
features.  A factor of 10 increase in exposure would permit line
profile diagnostics in a single line (\eli{Ne}{10}) and greatly
facilitate modeling.

\acknowledgments

This research was supported by NASA grant G03-4005A and by NASA
through the Smithsonian Astrophysical Observatory (SAO) contract
SV3-73016 for the Chandra X-Ray Center and Science Instruments.

{\it Facilities:} \facility{CXO(HETG)} \facility{CXO(LETG)}.

% for submission, comment out the next two lines and
% insert ms??.bbl (where ?? is the 2-digit version counter)
%
%\clearpage
\bibliographystyle{jwapjbib}                     % comment for submission
\bibliography{mnemonic,jw_abbrv,apj_abbrv,rsc}   % comment for submission

% \input{ms.bbl}                       % uncomment for submission
%% \begin{thebibliography}{}            

%\clearpage\input{tbl_linelist} 
\clearpage% aastex.cls definition uses \small for the ion font; 
% use local def
%%%%%%%%%%%%%%%
\begin{deluxetable}{rclcclllrrrr} 
\tabletypesize{\scriptsize}
\tablecaption{Line Measurements \label{tbl:linelist} }
\tablewidth{0pt}
\tablehead{
  \colhead{Mnemonic\tablenotemark{a}} & 
  \colhead{Use\tablenotemark{b}} & 
  \colhead{Ion} &
  \colhead{$\overline{\log T}$\tablenotemark{c}}&
  \colhead{$\lambda_t$\tablenotemark{d}}&
  \colhead{$\lambda_o$\tablenotemark{e}}&
  \colhead{$f_l$\tablenotemark{f}}&
  \colhead{$f_t$\tablenotemark{g}}&
  \colhead{$\delta f$\tablenotemark{h}}&
  \colhead{$\delta\chi$\tablenotemark{i}}
}
\startdata
Fe25HeLa&    E& \eli{Fe}{25}      & 7.81&  1.861&  1.867 (4.5) & 3.10 (1.88)   & 2.24  & 0.86  & 0.5\\
Ar17HeLar&    -& \eli{Ar}{17}     & 7.36&  3.949&  3.948 (15.0)& 0.34 (0.54)   & 2.12  & -1.78 & -3.3\\
Ar17HeLai&    -& \eli{Ar}{17}     & 7.31&  3.968&  3.970 (3.5) & 1.75 (0.79)   & 0.60  & 1.15  & 1.5\\
S16HLbB&    -& \eli{S}{16}        & 7.53&  3.992&  3.995 (8.6) & 0.84 (0.71)   & 0.50  & 0.34  & 0.5\\
S16HLa&    E& \eli{S}{16}         & 7.57&  4.730&  4.730 (6.7) & 2.57 (0.86)   & 2.31  & 0.26  & 0.3\\
S15HeLar&    E& \eli{S}{15}       & 7.20&  5.039&  5.040 (2.5) & 4.40 (1.20)   & 4.06  & 0.34  & 0.3\\
S15HeLai&    E& \eli{S}{15}       & 7.16&  5.065&  5.072 (5.6) & 1.47 (1.05)   & 0.98  & 0.49  & 0.5\\
S15HeLaf&    E& \eli{S}{15}       & 7.17&  5.101&  5.106 (6.0) & 2.95 (1.20)   & 1.43  & 1.52  & 1.3\\
Si13HeLb&    E& \eli{Si}{13}      & 7.06&  5.681&  5.688 (5.5) & 1.90 (1.10)   & 1.61  & 0.28  & 0.3\\
Si14HLa&    EHM& \eli{Si}{14}     & 7.39&  6.183&  6.183 (1.0) & 9.00 (0.83)   & 8.72  & 0.28  & 0.3\\
Si13HeLar&    EH& \eli{Si}{13}    & 7.03&  6.648&  6.647 (0.7) & 14.72 (1.00)  & 13.71 & 1.01  & 1.0\\
Si13HeLai&    -& \eli{Si}{13}     & 6.99&  6.687&  6.687 (2.5) & 2.91 (0.59)   & 2.70  & 0.21  & 0.4\\
Si13HeLaf&    -H& \eli{Si}{13}    & 7.01&  6.740&  6.739 (0.8) & 8.58 (0.75)   & 5.73  & 2.85  & 3.8\\
Mg12HLb&    E& \eli{Mg}{12}       & 7.22&  7.106&  7.107 (2.7) & 2.46 (0.46)   & 2.91  & -0.45 & -1.0\\
Al13HLaB&    -& \eli{Al}{13}      & 7.36&  7.171&  7.171 (1.7) & 3.03 (0.50)   & 2.07  & 0.96  & 1.9\\
Al12HeLar&    E& \eli{Al}{12}     & 6.98&  7.757&  7.758 (2.2) & 3.30 (0.64)   & 0.99  & 2.30  & 3.6\\
Al12HeLai&    -& \eli{Al}{12}     & 6.89&  7.805&  7.804 (7.1) & 0.59 (0.49)   & 0.35  & 0.24  & 0.5\\
Mg11HeLb&    E& \eli{Mg}{11}      & 6.87&  7.850&  7.853 (2.5) & 3.19 (0.73)   & 2.77  & 0.42  & 0.6\\
Al12HeLaf&    E& \eli{Al}{12}     & 6.89&  7.872&  7.872 (---) & 0.34 (0.40)   & 0.96  & -0.62 & -1.6\\
Fe23w7.90&    E& \eli{Fe}{23}     & 7.24&  7.901&  7.899 (3.7) & 2.77 (0.63)   & 0.33  & 2.44  & 3.9\\
Fe24w7.986&    -& \eli{Fe}{24}    & 7.43&  7.986&  7.972 (3.4) & 1.39 (0.68)   & 0.97  & 0.41  & 0.6\\
Fe24w7.996&    E& \eli{Fe}{24}    & 7.43&  7.996&  7.992 (5.0) & 1.72 (0.64)   & 0.50  & 1.22  & 1.9\\
Fe24w8.28&    E& \eli{Fe}{24}     & 7.40&  8.285&  8.285 (---) & 0.58 (0.56)   & 0.20  & 0.38  & 0.7\\
Fe23w8.30&    E& \eli{Fe}{23}     & 7.25&  8.304&  8.302 (4.0) & 1.66 (0.59)   & 1.20  & 0.45  & 0.8\\
Fe24w8.32&    E& \eli{Fe}{24}     & 7.42&  8.316&  8.316 (---) & 1.38 (0.57)   & 1.09  & 0.29  & 0.5\\
Fe24w8.38&    -& \eli{Fe}{24}     & 7.40&  8.376&  8.369 (---) & 1.39 (0.51)   & 0.43  & 0.96  & 1.9\\
Mg12HLa&    EHM& \eli{Mg}{12}     & 7.19&  8.422&  8.421 (0.6) & 24.46 (1.41)  & 22.53 & 1.93  & 1.4\\
Fe21w8.57&    E& \eli{Fe}{21}     & 7.06&  8.574&  8.574 (2.9) & 3.13 (0.85)   & 1.05  & 2.08  & 2.5\\
Fe23w8.81&    E& \eli{Fe}{23}     & 7.23&  8.815&  8.815 (---) & 1.94 (0.66)   & 1.26  & 0.68  & 1.0\\
Fe22w8.97&    -& \eli{Fe}{22}     & 7.13&  8.975&  8.983 (6.8) & 3.22 (0.73)   & 1.57  & 1.65  & 2.3\\
Mg11HeLar&    EHM& \eli{Mg}{11}   & 6.83&  9.169&  9.169 (0.7) & 23.78 (1.50)  & 23.14 & 0.64  & 0.4\\
Fe21w9.19&    -& \eli{Fe}{21}     & 7.07&  9.194&  9.200 (3.3) & 4.17 (0.86)   & 0.78  & 3.39  & 3.9\\
Mg11HeLai&    -& \eli{Mg}{11}     & 6.80&  9.230&  9.230 (4.1) & 4.72 (0.82)   & 3.77  & 0.95  & 1.2\\
Mg11HeLaf&    -HM& \eli{Mg}{11}   & 6.81&  9.314&  9.314 (0.9) & 12.81 (1.22)  & 11.04 & 1.77  & 1.5\\
Ne10HLd&    E& \eli{Ne}{10}       & 7.01&  9.481&  9.479 (1.8) & 5.68 (0.86)   & 3.12  & 2.56  & 3.0\\
Fe19w9.69&    -& \eli{Fe}{19}     & 6.93&  9.695&  9.681 (2.6) & 2.43 (0.67)   & 1.00  & 1.43  & 2.1\\
Ne10HLg&    EHM& \eli{Ne}{10}     & 7.00&  9.708&  9.708 (1.3) & 9.18 (1.00)   & 7.10  & 2.07  & 2.1\\
Fe20w9.73&    -& \eli{Fe}{20}     & 7.00&  9.727&  9.742 (2.6) & 0.93 (0.62)   & 0.96  & -0.03 & -0.1\\
Ni19w10.11&    E& \eli{Ni}{19}    & 6.84& 10.110& 10.108 (2.8) & 3.67 (0.84)   & 0.86  & 2.82  & 3.4\\
Fe20w10.12&    -& \eli{Fe}{20}    & 6.99& 10.120& 10.134 (---) & 1.09 (0.63)   & 1.08  & 0.02  & 0.0\\
Ne10HLb&    EHM& \eli{Ne}{10}     & 6.99& 10.239& 10.239 (0.8) & 19.50 (1.45)  & 22.77 & -3.27 & -2.3\\
Fe23w10.98&    E& \eli{Fe}{23}    & 7.24& 10.981& 10.978 (2.3) & 5.78 (1.22)   & 6.55  & -0.77 & -0.6\\
Ne9HeLg&    E& \eli{Ne}{9}        & 6.66& 11.001& 10.997 (5.0) & 4.39 (1.24)   & 4.02  & 0.38  & 0.3\\
Fe23w11.02&    -& \eli{Fe}{23}    & 7.24& 11.019& 11.020 (2.4) & 9.57 (1.42)   & 4.31  & 5.26  & 3.7\\
Fe24w11.03&    E& \eli{Fe}{24}    & 7.38& 11.029& 11.041 (4.0) & 4.01 (1.24)   & 4.75  & -0.74 & -0.6\\
Fe17w11.13&    E& \eli{Fe}{17}    & 6.74& 11.131& 11.133 (2.2) & 4.85 (1.09)   & 6.37  & -1.51 & -1.4\\
Fe24w11.18&    E& \eli{Fe}{24}    & 7.38& 11.176& 11.174 (1.6) & 10.20 (1.35)  & 8.59  & 1.61  & 1.2\\
Fe18w11.33&    E& \eli{Fe}{18}    & 6.85& 11.326& 11.325 (1.6) & 10.35 (1.44)  & 6.54  & 3.81  & 2.6\\
Fe18w11.53&    E& \eli{Fe}{18}    & 6.85& 11.527& 11.522 (3.4) & 6.77 (1.73)   & 4.50  & 2.27  & 1.3\\
Ne9HeLb&    E& \eli{Ne}{9}        & 6.64& 11.544& 11.545 (2.6) & 16.76 (1.82)  & 12.81 & 3.95  & 2.2\\
Fe23w11.74&    E& \eli{Fe}{23}    & 7.22& 11.736& 11.741 (1.8) & 16.89 (1.91)  & 14.17 & 2.72  & 1.4\\
Fe22w11.77&    E& \eli{Fe}{22}    & 7.13& 11.770& 11.772 (1.9) & 17.33 (2.03)  & 14.85 & 2.48  & 1.2\\
Ne10HLa&    EHM& \eli{Ne}{10}     & 6.95& 12.135& 12.133 (0.3) & 180.70 (5.71) & 175.97& 4.73  & 0.8\\
Fe23w12.16&    -& \eli{Fe}{23}    & 7.22& 12.161& 12.176 (1.1) & 5.99 (1.61)   & 7.95  & -1.96 & -1.2\\
Fe17w12.27&    -& \eli{Fe}{17}    & 6.73& 12.266& 12.261 (2.6) & 11.38 (2.26)  & 22.54 & -11.16& -4.9\\
Fe21w12.28&    EHM& \eli{Fe}{21}  & 7.05& 12.284& 12.286 (1.0) & 30.82 (2.84)  & 34.78 & -3.96 & -1.4\\
Fe20w13.38&    -t& \eli{Fe}{20}   & 6.99& 13.385& 13.370 (2.3) & 9.51 (2.14)   & 5.52  & 4.00  & 1.9\\
Fe19w13.42&    Et& \eli{Fe}{19}   & 6.92& 13.423& 13.423 (---) & 4.22 (3.53)   & 3.51  & 0.70  & 0.2\\
Ne9HeLar&    EHM& \eli{Ne}{9}     & 6.61& 13.447& 13.446 (1.4) & 104.60 (5.02) & 103.70& 0.90  & 0.2\\
Fe19w13.46&    Et& \eli{Fe}{19}   & 6.92& 13.462& 13.462 (---) & 6.08 (4.70)   & 7.98  & -1.89 & -0.4\\
Fe19w13.50&    Et& \eli{Fe}{19}   & 6.92& 13.497& 13.497 (---) & 17.00 (3.51)  & 14.00 & 3.00  & 0.9\\
Fe19w13.52&    Et& \eli{Fe}{19}   & 6.92& 13.518& 13.515 (1.8) & 39.88 (4.36)  & 30.86 & 9.02  & 2.1\\
Ne9HeLai\tablenotemark{j}&-& \eli{Ne}{9}
                                  & 6.58& 13.552& 13.552 (2.8) & 29.57 (3.54)  & 16.02 & 13.55 & 3.8\\
Fe19w13.64&    Et& \eli{Fe}{19}   & 6.92& 13.645& 13.639 (8.3) & 8.54 (2.41)   & 4.94  & 3.60  & 1.5\\
Ne9HeLaf&     HM& \eli{Ne}{9}     & 6.59& 13.699& 13.699 (1.2) & 61.51 (4.97)  & 51.80 & 9.71  & 2.0\\
Fe20w13.77&    -t& \eli{Fe}{20}   & 6.99& 13.767& 13.770 (3.1) & 16.66 (3.07)  & 4.70  & 11.96 & 3.9\\
Fe19w13.80&    -t& \eli{Fe}{19}   & 6.92& 13.795& 13.805 (3.1) & 15.70 (2.98)  & 12.35 & 3.35  & 1.1\\
Fe17w13.82&    -t& \eli{Fe}{17}   & 6.74& 13.825& 13.833 (2.6) & 19.22 (3.15)  & 17.50 & 1.72  & 0.5\\
Fe18w14.21&    EHM& \eli{Fe}{18}  & 6.84& 14.208& 14.202 (1.4) & 68.89 (5.49)  & 76.59 & -7.70 & -1.4\\
Fe18w14.26&    E& \eli{Fe}{18}    & 6.84& 14.256& 14.246 (5.5) & 15.39 (3.97)  & 14.73 & 0.66  & 0.2\\
Fe20w14.27&    E& \eli{Fe}{20}    & 6.98& 14.267& 14.269 (6.3) & 12.91 (3.43)  & 7.70  & 5.21  & 1.5\\
Fe18w14.34&    E& \eli{Fe}{18}    & 6.84& 14.343& 14.341 (6.4) & 11.11 (3.55)  & 9.10  & 2.01  & 0.6\\
Fe18w14.37&    E& \eli{Fe}{18}    & 6.84& 14.373& 14.376 (3.6) & 22.49 (4.08)  & 19.52 & 2.97  & 0.7\\
Fe18w14.53&    -HM& \eli{Fe}{18}  & 6.84& 14.534& 14.542 (2.4) & 24.93 (---)   & 14.83 & 10.10 & ---\\
Fe19w14.66&    E& \eli{Fe}{19}    & 6.92& 14.664& 14.670 (5.6) & 15.86 (3.53)  & 9.31  & 6.55  & 1.9\\
O8HLd&    E& \eli{O}{8}           & 6.74& 14.821& 14.821 (8.7) & 9.56 (3.07)   & 8.08  & 1.47  & 0.5\\
Fe17w15.01&    -HM& \eli{Fe}{17}  & 6.71& 15.014& 15.011 (0.7) & 148.70 (9.49) & 236.47& -87.77& -9.2\\
Fe19w15.08&    E& \eli{Fe}{19}    & 6.91& 15.079& 15.068 (4.2) & 21.43 (4.28)  & 10.48 & 10.95 & 2.6\\
O8HLg&    EHM& \eli{O}{8}         & 6.73& 15.176& 15.175 (2.4) & 28.34 (4.33)  & 18.32 & 10.02 & 2.3\\
Fe19w15.20&    E& \eli{Fe}{19}    & 6.92& 15.198& 15.202 (3.9) & 18.81 (3.80)  & 8.60  & 10.21 & 2.7\\
Fe17w15.26&    EHM& \eli{Fe}{17}  & 6.71& 15.261& 15.259 (1.1) & 74.11 (6.65)  & 66.83 & 7.28  & 1.1\\
Fe17w15.45&    E& \eli{Fe}{17}    & 6.70& 15.453& 15.453 (4.2) & 16.77 (3.31)  & 8.61  & 8.16  & 2.5\\
Fe18w15.62&    EM& \eli{Fe}{18}   & 6.83& 15.625& 15.625 (2.6) & 26.75 (4.29)  & 20.25 & 6.50  & 1.5\\
Fe18w15.82&    E& \eli{Fe}{18}    & 6.83& 15.824& 15.824 (3.2) & 16.17 (3.39)  & 12.38 & 3.79  & 1.1\\
Fe18w15.87&    E& \eli{Fe}{18}    & 6.83& 15.870& 15.870 (3.8) & 16.17 (3.86)  & 6.57  & 9.60  & 2.5\\
O8HLbB\tablenotemark{k}&-M& \eli{O}{8}
                                  & 6.71& 16.006& 16.005 (1.5) & 77.16 (7.96)  & 58.41 & 18.75 & 2.4\\
Fe18w16.07&    -M& \eli{Fe}{18}   & 6.83& 16.071& 16.072 (1.9) & 54.65 (6.77)  & 28.36 & 26.29 & 3.9\\
Fe19w16.11&    E& \eli{Fe}{19}    & 6.92& 16.110& 16.117 (8.8) & 10.30 (4.41)  & 13.65 & -3.35 & -0.8\\
Fe18w16.16&    E& \eli{Fe}{18}    & 6.83& 16.159& 16.159 (5.5) & 13.63 (4.76)  & 11.30 & 2.33  & 0.5\\
Fe17w16.78&    EM& \eli{Fe}{17}   & 6.71& 16.780& 16.772 (1.2) & 106.30 (9.32) & 107.63& -1.33 & -0.1\\
Fe17w17.05&    EM& \eli{Fe}{17}   & 6.71& 17.051& 17.048 (1.3) & 125.30 (10.79)& 128.13& -2.83 & -0.3\\
Fe17w17.10&    E& \eli{Fe}{17}    & 6.70& 17.096& 17.092 (1.1) & 133.70 (11.20)& 119.46& 14.24 & 1.3\\
Fe18w17.62&    E& \eli{Fe}{18}    & 6.83& 17.623& 17.618 (5.9) & 21.98 (5.51)  & 20.35 & 1.63  & 0.3\\
O7HeLb&    E& \eli{O}{7}          & 6.37& 18.627& 18.627 (7.2) & 18.04 (5.75)  & 13.66 & 4.38  & 0.8\\
O8HLa&    EM& \eli{O}{8}          & 6.66& 18.970& 18.967 (0.8) & 456.00 (26.06)& 439.97& 16.03 & 0.6\\
N7HLb&    -& \eli{N}{7}           & 6.53& 20.910& 20.925 (9.9) & 19.66 (8.64)  & 7.61  & 12.05 & 1.4\\
O7HeLar&    E& \eli{O}{7}         & 6.35& 21.602& 21.607 (3.8) & 98.15 (18.26) & 106.51& -8.36 & -0.5\\
O7HeLai&    -& \eli{O}{7}         & 6.32& 21.802& 21.799 (9.8) & 38.19 (15.20) & 13.50 & 24.69 & 1.6\\
O7HeLaf&    -& \eli{O}{7}         & 6.32& 22.098& 22.099 (6.1) & 74.15 (23.12) & 56.00 & 18.15 & 0.8\\
N7HLa&    E& \eli{N}{7}           & 6.48& 24.782& 24.788 (6.6) & 68.02 (17.33) & 59.21 & 8.81  & 0.5\\
\tableline
\enddata
\tablecomments{Values given in parentheses are the one standard
  deviation uncertainties on the preceding quantity. If the
  uncertainty has a value of ``---'', then either the confidence did
  not converge, or the parameter was frozen. ``Unused'' features were
  fit in order to obtain a good fit to a region and determine values
  for nearby interesting lines.}

\tablenotetext{a}{The mnemonic is a convenience for uniquely naming
  each feature.  It is comprised of the element and ion (in Arabic
  numerals) followed by a string indicating a wavelength and the
  wavelength (e.g., w16.78), or a code for the hydrogen-like (``H'') and
  helium-like ``He'' series, ``L'' for Lyman transition, one of ``a'',
  ``b'', ``g'', ``d'', or ``e'' for
  series lines $\alpha,\,\beta,\,\gamma,\,\delta,\,\epsilon$, and ``r'', ``i'',
  or ``f'' for resonance, intersystem, or forbidden lines. }

\tablenotetext{b}{``Use'' indicates whether the line was used in the
  emission measure reconstruction (``-'' for no, ``E'' for yes); 
  whether the feature was used in the composite line profile (``H''
  and ``M'' for \heg\ and \meg); or whether the line was used in the
  \eli{Ne}{9} triplet fit with the character, ``t''.}

\tablenotetext{c}{Average logarithmic temperature [Kelvins] of
  formation, defined as the first moment of the emissivity distribution.}

\tablenotetext{d}{Theoretical wavelengths of identification (from APED), in
  \AA. If the line is a multiplet, we give the wavelength of the
  strongest component.}

\tablenotetext{e}{Measured wavelength, in \AA\  (uncertainty
  is in m\AA).}

\tablenotetext{f}{Emitted source line flux is $10^{-6}$ times the tabulated value in 
  $[\mathrm{phot\,cm\mtwo\,s\mone}]$.}

\tablenotetext{g}{Model line flux is $10^{-6}$ times the tabulated value in 
  $[\mathrm{phot\,cm\mtwo\,s\mone}]$.}

\tablenotetext{h}{{Line} flux residual, $\delta f = f_o - f_t$.}

\tablenotetext{i}{$\delta\chi = (f_o - f_t)/\sigma_o$.}

\tablenotetext{j}{The \eli{Ne}{9} intercombination line flux is
  blended with \eli{Fe}{19} 13.551 \AA, and with \eli{Fe}{20} 13.535,
  13.558, 13.565 \AA\ lines.  The tabulated flux includes these
  blends.  About 85\% of the flux is from neon, as determined by the
  neighboring lines.}

\tablenotetext{k}{The \eli{O}{8} Lyman beta-like line is blended with
  \eli{Fe}{18} 16.004 \AA, whose flux is about 1.17 times that of
  \eli{Fe}{18} $\lambda\,15.62$, according to our \dem\ and \aped\ 
  model.  Hence, we estimate that the actual O8HLb flux is 45.9 (8.6).
}

\end{deluxetable}

\end{document}